\documentclass[manuscript,screen,nonacm]{acmart}

\setcopyright{acmlicensed}
\copyrightyear{2025}
\acmYear{2025}
\acmDOI{XXXXXXX.XXXXXXX}

\usepackage{tikz}
\usepackage{amsmath}

\usepackage{bbm}

\usepackage{array}
\usepackage{hyperref}
\usepackage{cleveref}
\usepackage{pifont}
\usepackage{xspace}
\usepackage{graphicx}
\usepackage{dsfont}
\usepackage{multirow}
\usepackage{subcaption}

\usepackage{svg}
\usepackage{todonotes}
\usepackage{subcaption}
\usepackage{booktabs}
\usepackage{wasysym}

\usepackage{url}

\usepackage{balance}

\newcommand{\unknown}{N/A\xspace}%
\newcommand{\labels}{\text{\ding{66}}}%
\newcommand{\probabilities}{\text{\ding{91}}}%
\newcommand{\explanations}{\text{\ding{107}}}%

\newcommand{\different}{$\blacksquare$}

\newcommand{\same}{$\square$}

\newcommand{\volsymbol}[1]{%
    \begin{tikzpicture}[scale=0.15, baseline=-0.5ex]
        \draw[line width=1mm] (0,0) -- (0,1.0); 
        \ifnum#1>1
            \draw[line width=1mm] (0.8,0) -- (0.8,1.0); 
        \fi
        \ifnum#1>2
            \draw[line width=1mm] (1.6,0) -- (1.6,1.0); 
        \fi
        \ifnum#1>3
            \draw[line width=1mm] (2.4,0) -- (2.4,1.0); 
        \fi
    \end{tikzpicture}%
}

\newcommand{\footurl}[1]{\footnote{\url{#1}}}%

\begin{document}

\title{I Stolenly Swear That I Am Up to (No) Good: Design and Evaluation of Model Stealing Attacks}

\author{Daryna Oliynyk}
\email{daryna.oliynyk@univie.ac.at}
\affiliation{%
  \institution{Christian Doppler Laboratory for Assurance and Transparency in Software Protection, University of Vienna}
  \city{Vienna}
  \country{Austria}
}

\author{Rudolf Mayer}
\email{rmayer@sba-research.org}
\affiliation{%
 \institution{SBA Research \& TU Wien}
 \city{Vienna}
 \country{Austria}}

\author{Kathrin Grosse}
\email{kathrin.grosse1@ibm.com}
\affiliation{%
 \institution{IBM Research Zurich}
 \city{Zurich}
 \country{Switzerland}}

\author{Andreas Rauber}
\email{rauber@ifs.tuwien.ac.at}
\affiliation{%
  \institution{Institute of Information Systems Engineering, TU Wien}
  \city{Vienna}
  \country{Austria}}

\renewcommand{\shortauthors}{Oliynyk et al.}

\begin{abstract}
Model stealing attacks endanger the confidentiality of machine learning models offered as a service. Although these models are kept secret, a malicious party can query a model to label data samples and train their own substitute model, violating intellectual property. While novel attacks in the field are continually being published, their design and evaluations are not standardised, making it challenging to compare prior works and assess progress in the field. 
This paper is the first to address this gap by providing recommendations for designing and evaluating model stealing attacks. To this end, we study the largest group of attacks that rely on training a substitute model -- those attacking image classification models. We propose the first comprehensive threat model and develop a framework for attack comparison. Further, we analyse attack setups from related works to understand which tasks and models have been studied the most. Based on our findings, we present best practices for attack development before, during, and beyond experiments and derive an extensive list of open research questions regarding the evaluation of model stealing attacks. Our findings and recommendations also transfer to other problem domains, hence establishing the first generic evaluation methodology for model stealing attacks.
\end{abstract}



\maketitle

\section{Introduction}
With the rising popularity of machine learning and its increasing deployment within real-world, practical applications and its ensuing exposure to end-users, the security aspects of machine-learning-based systems have gained wider attention. Several works showed that all machine learning models are vulnerable to attacks on their confidentiality, integrity, and availability, e.g., by evasion attacks in the form of adversarial examples~\cite{szegedy_intriguing_2014,biggio_wild_2018}.

In recent years, aspects of the intellectual property (IP) of machine learning models and systems exploiting those models have attracted more attention~\cite{oliynyk_i_2023} and are a concern of practitioners~\cite{grosse_machine_2023}. Especially relevant here is that under current legislation, machine learning models are protected as trade secrets~\cite{foss-solbrekk_three_2021}. In other words, if they are leaked, their ownership cannot be protected by legal means.
At the same time, machine learning models shared as black-box APIs are vulnerable to attacks at inference time, like model stealing (model extraction) attacks~\cite{tramer_stealing_2016}. While attacks have been investigated for different stealing goals against a vast range of machine learning tasks~\cite{oliynyk_i_2023}, ensuring comparability of attacks following the same objective became more crucial. 
Currently, the largest attack group, consisting of more than 40 works, is focused on stealing the behaviour of image classifiers through training a substitute model with similar functionality. However, the exact attack carried out can differ significantly, depending on the attacker's knowledge and capabilities -- which in turn leads to different levels of attack impact and difficulty. 
Consequently, a fair attack comparison becomes more challenging, as comparing only a quantitative performance measure of attacks is not enough. Moreover, as different metrics have been proposed to quantify the success rate of model stealing attacks (e.g.~\cite{jagielski_high_2020}), there is no unified way of performance reporting, and many works in the field report only an insufficient subset of relevant measures. This inability to objectively evaluate an attack leads to ambiguity in defining the current state-of-the-art and slows down both the research on attacks and, more crucially, countermeasure development.

Benchmarking and evaluation are important aspects in many related fields, such as information retrieval~\cite{armstrong_improvements_2009}, machine learning for classification problems~\cite{fernandez-delgado_we_2014}, fairness in computer vision~\cite{gustafson_facet_2023}, or robustness of machine learning models~\cite{rauber_foolbox_2017, croce_robustbench_2021}.
In the area of model stealing, we lack well-established practices for attack evaluation, and more crucially, we lack an understanding of which current practices are wrong or insufficient.

To the best of our knowledge, the evaluation methodology of model stealing attacks has not been addressed before.
With this paper, we therefore want to start a discussion about the criteria and recommendations that should be followed to ensure a fair, comprehensive, and comparable evaluation of model stealing attacks. For that purpose, we analyse substitute training attacks on image classifiers, which is the most actively developed sub-area of model stealing attacks. 

Based on our analysis, we provide a set of recommendations for future work in the field to be considered when evaluating and comparing new approaches.
Our main contributions are:
\begin{itemize}
    \item A comprehensive threat model on possible levels of attackers' knowledge, capabilities, and goals
    \item A systematisation of works on stealing image classification models along the proposed threat model
    \item The first framework for comparison of model stealing attacks that is based on the threat model
    \item An outlook of the current research status of the field based on the proposed framework, clearly demonstrating that (i) only a small fraction of prior attacks can be compared with each other, and (ii) several attack configuration have not been studied at all and a quarter of configurations have only been studied in one or two works
    \item An exhaustive analysis of previous works' experiment setups in terms of learning tasks, model architectures, and queries that define conditions and limitations of the performed evaluation
    \item Best practices that incorporate recommendations for implementation before, during, and beyond experiments to ensure transparent and comprehensive attack evaluation 
    \item An extensive list of current open questions related to the design and evaluation methodology of model stealing attacks  
\end{itemize}

The remainder of this paper is organised as follows. \Cref{sec:background} introduces terminology relevant for this work and discusses two similar areas, knowledge distillation and black-box adversarial attacks. Subsequently, in \Cref{sec:threat_model}, we present a comprehensive threat model for model stealing attacks on image classifiers, and present challenges we encountered while deriving it. Further, we address comparability of attacks in \Cref{sec:framework} by introducing our framework for selecting compatible prior attacks. In \Cref{sec:attack_setup}, we analyse frequent attack setups in terms of classification tasks, model architectures, and query patterns.
We then distil all prior observations into best practices in \Cref{sec:best_practices} and open research questions in \Cref{sec:open_research_questions}.
We further discuss how our findings transfer to other problem domains in \Cref{sec:discussion}.
Finally, \Cref{sec:related_work} gives an overview of related work in the field of evaluation methodology and \Cref{sec:conclusion} concludes the paper.

\section{Background}
\label{sec:background}
We begin with an overview of model stealing attacks, introducing the necessary terminology and notation. Further, we discuss two concepts that, while having different goals than model stealing, in some scenarios employ similar techniques -- knowledge distillation and black-box adversarial example attacks.

\subsection{Model Stealing}
Model stealing is an attack on the confidentiality of a machine learning model that aims at (partially) cloning the model by approximating its behaviour or obtaining information about its architecture, training hyperparameters, or learnable parameters (weights). In this work, we focus on attacks that aim to approximate the behaviour of image classification models. 

The model that an attacker tries to steal is called the \textit{target model}. We assume that the target model is exposed to third parties as-a-service, so that clients can send input \textit{queries} to the model and obtain corresponding outputs (predictions). Depending on the service functionality, model predictions can differ in granularity and expressiveness. For a target model $f$ that aims to classify every input query $x$ as one of $k$ classes $\{c_1, \ldots, c_k\}$, in this paper, we differentiate between three types of outputs. The first type is the prediction \textit{label} $y$ (also called hard label), which only displays the class $y=c \in \{c_1, \ldots, c_k\}$ to which query $x$ belongs according to the target model $f$, i.e. $f(x)=y$. The second type is prediction probabilities $(p_1, \ldots, p_k)$ (also confidence scores, soft labels or logits) that for each class $c_i$ show the probability $p_i$ of query $x$ belonging to class $c_i$. The third type of output is gradient-based explanations $e$, which are often returned in addition to the label $y$ or probabilities $(p_1, \ldots, p_k)$. Such explanations, for instance, Grad-CAM~\cite{selvaraju_grad_2020}, exploit gradients computed with respect to a particular layer (typically the last one) to display the most influential regions of the input image. 

To approximate the behaviour of the target model, the adversary can use the target model as an oracle to collect predictions for the \textit{attacker's data} $X$, which consists of individual queries $x_i \in X$. Depending on the prediction types defined above, the adversary might obtain labels $y_i \in Y$, probability scores $(p_1, \ldots, p_k)_i \in P$, or gradient-based explanations $e_i \in E$, typically returned together with labels or probabilities. By combining the attacker's data $X$ with corresponding predictions $Y$ (or $P$, $Y\cup E$, $ P\cup E$), the adversary can create their own dataset and train a so-called \textit{substitute model} $\hat{f}$ that aims to replicate the behaviour of the target model $f$.

The size of the attacker's dataset $|X|$, i.e., the number of queries the adversary sends to the target model, is a common measurement for assessing the efficiency of an attack. For assessing the effectiveness, three metrics are commonly used: accuracy, fidelity, and transferability. Accuracy shows the performance of the substitute model on the original learning task of the target model, for instance, by a test set $\{(x_i, y_i)\}_{i=1}^{N}$, i.e. 
\begin{equation}
Accuracy = \dfrac{1}{N} \sum_{i= 1}^{N}\mathbbm{1}(\hat{f}(x_i)=y_i).
\end{equation}
Fidelity shows similarity between the substitute and target model predictions: 
\begin{equation}
Fidelity = \dfrac{1}{N} \sum_{i= 1}^{N}\mathbbm{1}(\hat{f}(x_i) = f(x_i)).
\end{equation}
Finally, transferability shows how many of the adversarial examples crafted to fool the substitute model can also fool the target model into predicting the wrong class:
\begin{equation}
Transferability = \dfrac{1}{N} \sum_{i= 1}^{N}\mathbbm{1}(\hat{f}(x_i^{\prime})  \neq \hat{f}(x_i) \implies f(x^{\prime}_i) \neq f(x^{\prime}_i)),
\end{equation}
where $x_i^{\prime} = x_i +\varepsilon$ is an adversarial example for the substitute model $\hat{f}$, i.e. $\hat{f}(x_i^{\prime}) \neq \hat{f(x_i)}$ for a small $\varepsilon$.

While having only queries as an interaction channel with the target model is a common assumption for stealing models offered as a service, there is a whole group of model stealing attacks that assume side-channel access to the target model~\cite{oliynyk_i_2023}. This includes hardware, e.g. electromagnetic~\cite{batina_csi_2019} or power~\cite{xiang_open_2020} side-channels, and software side channels like time~\cite{duddu_stealing_2019} or cache~\cite{yan_cache_2020}. Side-channel attacks exploit leaked information to reveal the exact model properties, such as architecture or learned parameters. On account of (i) the stealing goal we consider (behaviour approximation) being different, and (ii) side-channel attacks assuming different attacker's capabilities that are not straightforwardly comparable with the query-based attacks, we exclude side-channel attacks from the scope of this work.

The diversity of data domains and tasks that are targeted with model stealing attacks keeps increasing. While analysing the field, we gathered 97 papers introducing methods for stealing the behaviour of machine learning models, almost half (47) of which focus on image classifiers, thus forming the most significant group of papers. Among other image-related tasks, attacks target encoders~\cite{liu_stolenencoder_2022}, and object detectors~\cite{liang_imitated_2022}. In the text domain, attacks have evolved from stealing simple recurrent neural networks for text classification~\cite{pal_framework_2019} to more complex BERT models~\cite{krishna_thieves_2020} and even large language models~\cite{gudibande_false_2024}. Graph data was also targeted in several works targeting graph neural networks~\cite{defazio_adversarial_2020, he_stealing_2021}. Besides, model stealing attacks also appeared in fields of recommender~\cite{yue_black-box_2021} and autonomous driving systems~\cite{zhang_play_2022}. However, studies outside of image classification are rather scattered at the moment and have not reached a sufficient level of development to draw overarching conclusions regarding the evaluation methodology.

\subsection{Knowledge Distillation}
In the following, we describe a concept similar to model stealing -- knowledge distillation. In particular, we discuss different categories of methods and which of them can be adapted to model stealing. 

Knowledge distillation (KD) is a model compression technique that aims to distil knowledge from a large deep neural network (the \textit{teacher}) into a simpler, smaller network (the \textit{student})~\cite{hinton_distilling_2015}. Although white-box access to a teacher and availability of original training data are commonly assumed for KD, various scenarios similar to model stealing have been proposed. A recent survey by Gou et al.~\cite{gou_knowledge_2021} classifies KD approaches based on (i) the knowledge a student obtains from the teacher, (ii) the distillation scheme, (iii) the teacher-student architecture, and (iv) the algorithm. Following their categorisation, KD techniques can be utilised for model stealing through substitute model training if they (i) are response-based, i.e., the student only obtains outputs of the teacher, and (ii) use offline distillation, i.e., the student is trained after the teacher.
The similarity between KD and model stealing was also spotted by Ma et al.~\cite{ma_undistillable_2021} and later utilised by two works to perform stealing attacks~\cite{kundu_analyzing_2021, jandial_distilling_2022}. 
Although some papers have been published on response-based offline knowledge distillation (also called black-box KD)~\cite{wang_zero_2021} and data-free knowledge distillation~\cite{wang_unpacking_2024}, exploring the suitability of KD methods for model stealing is out of the scope of this paper and is left for future work.

\subsection{Black-box Adversarial Example Attacks}
Another field that shares some similarities with model stealing is adversarial examples crafted under black-box access to the target model.
Adversarial examples~\cite{szegedy_intriguing_2014,goodfellow_explaining_2015} are minute perturbations to data samples that cause a model to change its prediction, e.g., to misclassify a sample to a different class. These attacks are primarily studied for the image domain but also apply to other data types. The aim of the perturbation is to be not (easily) human detectable while fooling the model with a high success chance. Numerous methods exist for crafting these examples, mainly requiring white-box access to the model to be attacked.

There are two groups of attacks that operate under black-box access. The first group trains a substitute model to simulate the decision boundary of the target model and relies on the transferability of adversarial examples~\cite{papernot_practical_2017}. The second group directly optimises adversarial examples by approximating gradients of the target model~\cite{chen_zoo_2017}. Below, we elaborate on mechanisms exploited by the first group, which has a similar methodology to model stealing. 

Papernot et al.~\cite{papernot_transferability_2016} were the first to demonstrate that one can employ model stealing to create a substitute model, and then craft adversarial examples for that stolen model with any white-box technique. If the two models are similar, adversarial examples crafted for the substitute model will also have a similar probability of succeeding in fooling the target model. Several further works focused on specifically optimising the training process for obtaining high transferability scores~\cite{zhou_dast_2020, chen_query_2024, sun_exploring_2022}.

While such black-box adversarial attacks have similar approaches to model stealing, they often can not be easily compared to approaches focused on model stealing itself, as they differ in purpose and evaluation. Many works only measure transferability, but not the other properties generally required for a successful model theft, like the performance of the stolen model on the original task of the target model.
Therefore, we do not consider works that only address transferability in our analysis. However, we do recommend evaluating the fidelity and accuracy of the substitute models trained with these methods, as they might constitute significant progress over current methods.

\section{Threat Model Overview} 
\label{sec:threat_model}

We start by analysing the threat models covered in $47$ relevant papers in \Cref{tab:threat_model}.
The threat models reported in that table are primarily derived by us from the experimental setup of the papers, as (i) only a minor fraction of the papers explicitly defined their threat models, and (ii) in some cases, the defined threat model was not followed and consequently verified in any of the experiments. While some of the approaches can potentially be applied under other conditions, one would need to verify this assumption with an empirical evaluation.

Based on the aspects of the threat model described in~\cite{biggio_wild_2018}, in the following, we characterise the attacker's knowledge, capabilities, and goals for substitute training attacks on image classifiers and how they are represented in \Cref{tab:threat_model}.

\begin{table}[ht!]
\centering
\caption{Threat models in prior works. Attacker's knowledge: data (first column), target outputs (column "Outputs": \protect\labels\ - labels, \protect\probabilities\ - confidence scores, \protect\explanations\ -  explanations or gradients), and target architecture (column "Model": \protect\different\ if target and substitute are different, \protect\same\  -- if the same). Attacker's capabilities: maximum number of queries (column "QB": \protect\volsymbol{1} - 10k, \protect\volsymbol{2} - 10k-100k, \protect\volsymbol{3} - 100k-1m, \protect\volsymbol{4} - 1m). Attacker's goal: evaluation metrics (column "Metrics":  A - accuracy, F - fidelity, T - transferability).}
\label{tab:threat_model}
\resizebox{0.55\linewidth}{!}{
\begin{tabular}{llcclc}
\toprule
                                                          & \textbf{Paper}                                         & \textbf{Outputs}       & \textbf{Model}   & \textbf{QB}   & \textbf{Metrics}                                  \\ \midrule
\multirow{17}{*}{Original data} & Pengcheng et al. \cite{pengcheng_query-efficient_2018} & -               & \different       & \volsymbol{2} & AF\textcolor{lightgray}{T}                        \\
                                                          & Papernot et al. \cite{papernot_practical_2017}         & \labels                & \different       & \volsymbol{1} & A\textcolor{lightgray}{F}T                        \\
                                                          & Pape et al. \cite{pape_limitations_2023}               & \labels                & \different/\same & -      & \textcolor{lightgray}{A}F\textcolor{lightgray}{T} \\
                                                          & Yan et al. \cite{yan_holistic_2023}                    & \labels                & \different/\same & \volsymbol{1} & AF\textcolor{lightgray}{T}                        \\
                                                          & Juuti et al. \cite{juuti_prada_2019}                   & \labels/\probabilities & \different/\same & \volsymbol{3} & \textcolor{lightgray}{A}FT                        \\
                                                          & Jagielski et al. \cite{jagielski_high_2020}            & \probabilities         & -         & \volsymbol{1} & AF\textcolor{lightgray}{T}                        \\
                                                          & Liu et al. \cite{liu_shrewdattack_2023}                & \probabilities         & -         & \volsymbol{2} & A\textcolor{lightgray}{FT}                        \\
                                                          & Yu et al. \cite{yu_cloudleak_2020}                     & \probabilities         & \different       & \volsymbol{1} & A\textcolor{lightgray}{FT}                        \\
                                                          & Zhao et al. \cite{zhao_extracting_2023}                & \probabilities         & \different       & \volsymbol{1} & \textcolor{lightgray}{A}FT                        \\
                                                          & Zhang et al. \cite{zhang_thief_2021}                   & \probabilities         & \different/\same & \volsymbol{1} & A\textcolor{lightgray}{FT}                        \\
                                                          & Chen et al. \cite{chen_d-dae_2023}                     & \probabilities         & \same            & \volsymbol{2} & AF\textcolor{lightgray}{T}                        \\
                                                          & He et al. \cite{he_drmi_2021}                          & \probabilities         & \same            & \volsymbol{3} & A\textcolor{lightgray}{F}T                        \\
                                                          & Li et al. \cite{li_model_2025}                       & \probabilities         & \same            & \volsymbol{3} & A\textcolor{lightgray}{FT}                        \\
                                                          & Liu et al. \cite{liu_ml-doctor_2022}                   & \probabilities         & \same            & \volsymbol{3} & AF\textcolor{lightgray}{T}                        \\
                                                          & Yan et al. \cite{yan_explanation_2023}                 & \explanations          & -         & \volsymbol{1} & \textcolor{lightgray}{A}F\textcolor{lightgray}{T} \\
                                                          & Yan et al. \cite{yan_towards_2022}                     & \explanations          & \different       & -      & A\textcolor{lightgray}{FT}                        \\
                                                          & Milli et al. \cite{milli_model_2019}                   & \explanations          & \different/\same & \volsymbol{1} & A\textcolor{lightgray}{FT}                        \\ \midrule
\multirow{4}{*}{PD data}        & Correia-Silva et al. \cite{correia-silva_copycat_2018} & \labels                & -         & \volsymbol{3} & A\textcolor{lightgray}{FT}                        \\
                                                          & Yu et al. \cite{yu_cloudleak_2020}                     & \probabilities         & \different       & \volsymbol{1} & A\textcolor{lightgray}{FT}                        \\
                                                          & Zhao et al. \cite{zhao_extracting_2023}                & \probabilities         & \different       & \volsymbol{1} & \textcolor{lightgray}{A}FT                        \\
                                                          & Xie et al. \cite{xie_game_2022}                        & \probabilities         & \different       & \volsymbol{1} & AF\textcolor{lightgray}{T}                        \\ \midrule
\multirow{17}{*}{nPD data}      & Wang et al. \cite{wang_enhance_2022}                   & \labels                & -         & \volsymbol{2} & A\textcolor{lightgray}{FT}                        \\
                                                          & Correia-Silva et al. \cite{correia-silva_copycat_2018} & \labels                & -         & \volsymbol{4} & A\textcolor{lightgray}{FT}                        \\
                                                          & Mosafi et al. \cite{mosafi_stealing_2019}              & \labels                & \different       & -      & A\textcolor{lightgray}{FT}                        \\
                                                          & Karmakar et al. \cite{karmakar_marich_2023}            & \labels                & \different       & \volsymbol{1} & AF\textcolor{lightgray}{T}                        \\
                                                          & Gong et al. \cite{gong_inversenet_2021}                & \labels                & \different       & \volsymbol{2} & AF\textcolor{lightgray}{T}                        \\
                                                          & Jindal et al. \cite{jindal_army_2024}           & \labels                & \different       & \volsymbol{2} & AFT                                               \\
                                                          & Yan et al. \cite{yan_holistic_2023}                    & \labels                & \different/\same & \volsymbol{1} & AF\textcolor{lightgray}{T}                        \\
                                                          & Wang et al. \cite{wang_black-box_2022}                 & \labels                & \different/\same & \volsymbol{2} & AFT                                               \\
                                                          & Yang et al. \cite{yang_swifttheft_2024}            & \labels                & \same            & \volsymbol{2} & \textcolor{lightgray}{A}F\textcolor{lightgray}{T} \\
                                                          & Orekondy et al. \cite{orekondy_knockoff_2019}          & \labels/\probabilities & \different/\same & \volsymbol{2} & A\textcolor{lightgray}{FT}                        \\
                                                          & Pal et al. \cite{pal_framework_2019, pal_activethief_2020}                   & \labels/\probabilities & \different/\same & \volsymbol{3} & \textcolor{lightgray}{A}F\textcolor{lightgray}{T} \\
                                                          & Atli et al. \cite{atli_extraction_2020}                & \labels/\probabilities & \same            & \volsymbol{3} & A\textcolor{lightgray}{FT}                        \\
                                                          & Zhao et al. \cite{zhao_extracting_2023}                & \probabilities         & \different       & \volsymbol{1} & \textcolor{lightgray}{A}FT                        \\
                                                          & Xie et al. \cite{xie_game_2022}                        & \probabilities         & \different       & \volsymbol{1} & AF\textcolor{lightgray}{T}                        \\
                                                          & Khaled et al. \cite{khaled_careful_2022}               & \probabilities         & \different       & \volsymbol{2} & AFT                                               \\
                                                          & Barbalau et al. \cite{barbalau_black-box_2020}         & \probabilities         & \different/\same & -      & A\textcolor{lightgray}{FT}                        \\
                                                          & Chen et al. \cite{chen_d-dae_2023}                     & \probabilities         & \same            & \volsymbol{2} & AF\textcolor{lightgray}{T}                        \\ \midrule
\multirow{15}{*}{Data-free}     & Yang et al. \cite{yang_efficient_2023}                 & \labels                & \different       & -      & A\textcolor{lightgray}{FT}                        \\
                                                          & Sanyal et al. \cite{sanyal_towards_2022}               & \labels                & \different       & \volsymbol{4} & A\textcolor{lightgray}{FT}                        \\
                                                          & Hondru et al. \cite{hondru_towards_2025}               & \labels/\probabilities & \different       & \volsymbol{2} & A\textcolor{lightgray}{FT}                        \\
                                                          & Zhang et al. \cite{zhang_towards_2022}                 & \labels/\probabilities & \different       & \volsymbol{3} & A\textcolor{lightgray}{F}T                        \\
                                                          & Hong et al. \cite{hong_exploring_2023}                 & \labels/\probabilities & \different       & \volsymbol{4} & A\textcolor{lightgray}{FT}                        \\
                                                          & Beetham et al. \cite{beetham_dual_2023}                & \labels/\probabilities & \different       & \volsymbol{4} & A\textcolor{lightgray}{FT}                        \\
                                                          & Yuan et al. \cite{yuan_es_2022}                        & \labels/\probabilities & \different/\same & -      & A\textcolor{lightgray}{F}T                        \\
                                                          & Lin et al. \cite{lin_quda_2023}                        & \labels/\probabilities & \different/\same & \volsymbol{3} & A\textcolor{lightgray}{FT}                        \\
                                                          & Rosenthal et al. \cite{rosenthal_disguide_2023}        & \labels/\probabilities & \different/\same & \volsymbol{4} & A\textcolor{lightgray}{FT}                        \\
                                                          & Truong et al. \cite{truong_data-free_2021}             & \probabilities         & \different       & \volsymbol{4} & A\textcolor{lightgray}{FT}                        \\
                                                          & Liu et al. \cite{liu_efficient_2024}                  & \probabilities         & \different       & \volsymbol{4} & AF\textcolor{lightgray}{T}                        \\
                                                          & Kariyappa et al. \cite{kariyappa_maze_2021}            & \probabilities         & \different       & \volsymbol{4} & A\textcolor{lightgray}{FT}                        \\
                                                          & Roberts et al. \cite{roberts_model_2019}               & \probabilities         & \same            & \volsymbol{3} & A\textcolor{lightgray}{FT}                        \\
                                                          & Miura et al. \cite{miura_megex_2021}                   & \explanations          & \different       & \volsymbol{4} & A\textcolor{lightgray}{FT}                        \\
                                                          & Yan et al. \cite{yan_explanation-based_2023}           & \explanations          & \different       & \volsymbol{4} & A\textcolor{lightgray}{FT}                
                                                          \\ \bottomrule       
\end{tabular}
}
\end{table}

\subsection{Attacker's Knowledge}
\label{sec:threat_model:knowledge}
The attacker's knowledge majorly defines the strength of an attack, consequently also characterising its difficulty.  
For model stealing, we identify three assets the attacker may have different knowledge of, namely the target model's training data, the target model's outputs, and the target model's architecture. Further, we argue that there is another component that was overlooked in previous work -- the usage of pre-trained models. 

\textbf{Target model's data.} 
In order to train a substitute model, an attacker has to create their own labelled dataset. This dataset usually starts with unlabelled data samples, which need to be labelled by the target model. However, the attacker's dataset can significantly differ in quality depending on the attacker's knowledge of the original classification task. 
We grouped prior works in \Cref{tab:threat_model} by the type of data available to the adversary:
(i) \textit{original} data that comes from the same distribution as the training data of the target model, 
(ii) \textit{problem-domain (PD)} data that has the same context as the original data but a different distribution,
(iii) \textit{non-problem-domain data (nPD)} that does not fit into the context or distribution of the original data,
and (iv) a \textit{data-free} case, which assumes that no real data was used to obtain the substitute model.

\textbf{(\labels/\probabilities/\explanations) Outputs of the target model.} Predictions of the target model can be disclosed to end-users with different degrees of detail. The column "Outputs" in \Cref{tab:threat_model} shows the outputs utilised by prior works, namely (i) (hard) labels (\labels), (ii) confidence scores or logits (\probabilities), (iii) both label-only and confidence score outputs were studied (\labels/\probabilities), (iv) explanations of predictions or gradients (\explanations), (iv) the outputs were not specified explicitly (-).

\textbf{ (\different/\same) Target model's architecture.} If an attacker knows the architecture of the target model, they might utilise the same architecture to train a substitute model. In \Cref{tab:threat_model}, we report this information under the column "Model" with four possible options: (i) the substitute model has the same architecture as the target (\same), (ii) a different one (\different), (iii) both previous options were studied (\different/\same), and (iv) not enough information was provided to conclude any of the above (-).

\textbf{Pre-trained models.} Another aspect of the attacker's knowledge concerns pre-trained models. The distribution of data used to pre-train weights of a model might have similar properties to the distribution of data from the target model's learning task. Then, by using this model as a starting point for training a substitute, an adversary can gain an advantage from the distribution similarity and achieve better performance than with a substitute trained from scratch~\cite{atli_extraction_2020, zhang_thief_2021}. For image classification, for instance, a common practice is to use models pre-trained on ImageNet as a starting point for training substitute models. 

Pre-trained auxiliary models like image-generating models or autoencoders can also be utilised to improve the performance of the attack. In particular, they can be used to initialise synthetic data generators for attacks with weak knowledge about the target model's original training data~\cite{barbalau_black-box_2020, lin_quda_2023, yang_continual_2023}. As these models are usually task-specific, one can not assume that they are equally available or useful for all classification tasks. Moreover, if we go beyond the scope of image classification, the availability of public pre-trained models differs depending on the data domain. Therefore, the availability of pre-trained models should be considered as one of the factors that impact the strength of an attack. 
The discussion of whether the availability of pre-trained models should be considered part of the attacker's knowledge was not addressed in previous works. Moreover, information about whether a substitute model is pre-trained is often omitted (see \Cref{sec:attack_setup:substitute_model} for more details), which makes the analysis of prior works impossible. Hence, we omit this information from \Cref{tab:threat_model}, but we encourage researchers to consider this in future evaluations.

\subsection{Attacker's Capabilities}
\label{sec:threat_model:capabilities}
It is a common assumption for substitute model training attacks that the target model is shared in a black-box manner (e.g., via an API) and that the only interaction possible is querying. Therefore, the query budget for conducting an attack is the main characteristic of the attacker's capabilities. 

\textbf{(\volsymbol{4}) Query budget.} We differentiate attacks on their query budget (QB) in \Cref{tab:threat_model}, defined as the maximum number of queries spent in the corresponding paper. We split query budgets into four groups: (i) below 10,000 queries (\volsymbol{1}), (ii) between 10,000 and 100,000 queries (\volsymbol{2}), (iii) between 100,000 and 1,000,000 queries (\volsymbol{3}), and (iv) more than 1,000,0000 queries (\volsymbol{4}). In some works, information about queries was not provided (-). 

\subsection{Attacker's Goal}
\label{sec:threat_model:attack_goal}
We consider model stealing attacks that aim to train a substitute model with a behaviour close to the target model's behaviour. Reaching this goal can be defined and quantified in two ways, accordingly to prior ways~\cite{jagielski_high_2020, oliynyk_i_2023}. In the following, we discuss both of them and explain how they are incorporated in this work. 

Jagielski et al. differentiate between three goals: functionally equivalent extraction, fidelity extraction, and task accuracy extraction~\cite{jagielski_high_2020}. 
Creating a functionally equivalent model means that the substitute model should give the same predictions as the target model on the whole input space. Fidelity extraction releases this requirement to a specific input data distribution, usually, the one that corresponds to the task the target model was designed to solve. Both goals are measured using the same metric, fidelity. Finally, task accuracy extraction means that the substitute model has high performance on the original target model's task, which is commonly measured using accuracy. 

Another goal categorisation for behaviour stealing is proposed by Oliynyk et al.~\cite{oliynyk_i_2023}. According to the authors, the substitute model aims to reach either the same level of effectiveness as the target model or prediction consistency with the target model. In the first case, the performance of the substitute is measured by accuracy. For prediction consistency, the prioritised metric depends on the data used for the evaluation. For real data, fidelity is measured, whereas for adversarial examples, it is transferability.

\textbf{(\textcolor{lightgray}{AFT}) Metrics.} We combine the suggestions from both prior works~\cite{jagielski_high_2020, oliynyk_i_2023} and consider accuracy (A), fidelity (F), and transferability (T) as indicators of the specific goals set in the reviewed papers in the last column of \Cref{tab:threat_model}.

\subsection{Limitations and Challenges}
\label{sec:threat_model:limitations_and_challanges}

To the best of our knowledge, we are the first to attempt a comprehensive analysis of model stealing threat models. Due to the lack of rigorous definitions, we encountered several challenges while classifying earlier attacks in \Cref{tab:threat_model}. In the following, we express these challenges in the form of open questions, outline our suggested solutions, and describe their limitations.
The way we resolve raised questions can only be taken as suggestions, as there are also reasonable arguments for other solutions. Primarily, we highlight those issues to bring more awareness to the lack of unified definitions and terminology.

\subsubsection{Attacker's Knowledge}
\label{sec:threat_model:limitations_and_challanges:knowledge}

While deriving assumptions about the attacker's knowledge from prior work, we encountered difficulties in defining the types of the attacker's data. Identifying if substitute and target architectures are the same or how detailed the outputs of the target models are primarily requires reporting corresponding information. Similarly, knowledge about pre-trained models can be communicated by explicitly reporting information about the utilised models. However, the body of related work is full of ambiguities when it comes to defining the type of data used. Below we list the most common controversies and how we mapped each of them into the categories in \Cref{tab:threat_model}. 

\textit{Original or problem-domain data?}
There are two positions on how original data can be defined. The first position considers any samples drawn from the distribution of the target model's training data as original data. The second position defines exclusively the training data of the target model as the original data. Therefore, if half of the training set of a dataset is used to train the target model, and the other half is used for the substitute model training, the data type can be classified as either original (the first position) or problem-domain (the second position).
We agree that this scenario differs from training on exactly the same data, but compared to what we consider problem-domain data, the distribution of data used is significantly closer to the target model's training data.
Hence, if an adversary uses part of the original dataset on which the target model was trained (even samples unseen by the target model) to train the substitute model, we classify its data type as original in \Cref{tab:threat_model}.

\textit{Problem-domain or non-problem-domain data?}
This question arises when a dataset that is supposed to be non-problem-domain data (nPD) has some contextual or sample intersection with the original data. We obtained two approaches to address this issue: (i) removing the same categories or samples from the attacker's dataset, and (ii) reporting quantified intersection between classes or datasets, while keeping the data unmodified. While keeping this data gives a benefit to the adversary, such intersections are often unintentional and might also occur in real-world scenarios. As in all works more than 50\% of data samples are indeed nPD, we classified all of them as ones using nPD data.  

\textit{Data-free or non-problem-domain data?}
Another ambiguity is over the definition of a data-free attack. We encounter two points of view: (i) a data-free attack does not use any real data at any stage of a model stealing attack, and (ii) a data-free attack does not use any real data to train a substitute model. 
In the second scenario, nPD data may have been used, for instance, to train a synthetic data generator that will subsequently produce PD-like data for the substitute model training. In both cases, the substitute model itself is trained without any real data, i.e., in a data-free manner. 
However, if an attack relies on having some nPD data to train a generator, then its data availability assumption does not differ from an attack that directly queries nPD data and trains a substitute model on it. On the other side, attacks that do not query nPD data often rely on open-source pre-trained generative models, implicitly using nPD data. For this reason, we categorised as data-free all attacks that do not use real data for training a substitute, regardless of the data used to train the generative model.

\subsubsection{Attacker's Capabilities}
\label{sec:threat_model:limitations_and_challanges:capabilities}

We represented attacker's capabilities in \Cref{tab:threat_model} as the maximum number of queries used by an attack. This results in two questions concerning queries: (i) which number we should actually consider, and (ii) whether counting queries is a representative metric for the attacker's capabilities and the efficiency of the attack. 

\textit{Which number should we consider?}
Our first point concerns two terms that are sometimes used interchangeably in the literature: query budget and the number of queries. In \Cref{tab:threat_model}, we reported query budgets, which we defined as the maximal number of queries used in a work. However, these numbers do not characterise how many queries an attack actually needs; it is an upper bound that the authors defined for their work. In fact, an attack performance might converge using fewer queries. This difference between the upper bound and the actually required amount of queries represents the difference between the query budget and the number of queries. While using the number of queries would reflect the strength of the attack better, deriving these numbers from related works remains challenging. We elaborate more on this question in \Cref{sec:open_research_questions:queries}.

\textit{Is counting queries enough?}
The absolute number of queries does not consider the difficulty of stealing. Such difficulty can be represented by the complexity of the target model (\cite{oliynyk_i_2023}) or by the complexity of its classification task. Therefore, a more explanatory metric would include the number of queries (or the query budget) in relation to the size of the target model or its training dataset. 
Another issue is that the number of queries (or query budget) allows comparison of attacks with each other, but is not a very interpretable metric for real-world scenarios. Instead, measuring the actual cost of training a substitute model would provide more insights into the attack efficiency. At the moment, it remains an open question how to actually estimate the price of the model. However, in terms of model stealing, one can estimate the cost of querying the target model by using available APIs, which can serve as a lower bound of the model cost.

\subsubsection{Attacker's Goal}
\label{sec:threat_model:limitations_and_challanges:goal}

Aligning prior works based on their goals is essential for their comparability. As we can see from \Cref{tab:threat_model}, no metric was reported in all papers, making a comparison of some methods complicated. We dedicate the whole following to the problem of comparability, whereas below we focus on another issue -- measuring transferability. 

\textit{How to measure transferability?} Transferability has no commonly agreed-upon method of measuring, unlike accuracy and fidelity. A major obstacle is that one needs to decide which adversarial example crafting technique to use and set its hyperparameters, which heavily influences the difficulty of achieving high transferability and -- it is generally easier to achieve high transferability scores on adversarial examples that contain larger perturbations. This, in turn, may influence the scores even more than the actual stealing method. Whereas we report transferability as one of the goals in \Cref{tab:threat_model}, the actual scores can thus not be used to compare the attacks. 
To illustrate this reasoning, we gathered information from 12 papers that reported transferability scores in \Cref{tab:transferability}. Whereas there are clearly two dominant approaches, namely the Fast Gradient Sign Method (FGSM)~\cite{goodfellow_explaining_2015} and Projected Gradient Descent (PGD)~\cite{madry_towards_2018}, the maximum perturbation value that regulates the strength of the perturbation varies from paper to paper, making the comparison unfair  

    \begin{table}[ht!]
    \centering
    \caption{Transferability measurement in related work with used attack and measured as perturbation size.}
    \label{tab:transferability}
    \resizebox{0.8\linewidth}{!}{
\begin{tabular}{lll}
\toprule
\textbf{Paper}                        & \textbf{Adversarial algorithm}                                & \textbf{Max. perturbation}        \\ \midrule
Beetham et al. \cite{beetham_dual_2023}              & \cite{goodfellow_explaining_2015} (FGSM), \cite{madry_towards_2018} (PGD)                                        & 0.01 \\
Jindal et al. \cite{jindal_army_2024}        & \cite{madry_towards_2018} (PGD)                                                                                  & 0.03 \\ 
Wang et al. \cite{wang_black-box_2022}            & \cite{madry_towards_2018} (PGD)                                                                                  & 0.03 \\ 
Khaled et al. \cite{khaled_careful_2022}            & \cite{goodfellow_explaining_2015} (FGSM), \cite{madry_towards_2018} (PGD)                                        & 0.03, 0.05, 0.10, 0.15                                                                    \\ 
Zhao et al. \cite{zhao_extracting_2023}           & \cite{goodfellow_explaining_2015} (FGSM)                                                                         & \begin{tabular}[l]{@{}l@{}}0.03, 0.06, 0.09, 0.12, 0.15, 0.18, 0.24\end{tabular}       \\ 
Zhang et al. \cite{zhang_towards_2022}             & \cite{goodfellow_explaining_2015} (FGSM), \cite{kurakin_adversarial_2017} (BIM), \cite{madry_towards_2018} (PGD) & 0.03, 0.13 \\ 
Yuan et al. \cite{yuan_es_2022}                   & \cite{madry_towards_2018} (PGD)                                                                                  & 0.03, 0.3                                                                         \\ 
Papernot et al. \cite{papernot_practical_2017}        & \cite{goodfellow_explaining_2015} (FGSM), \cite{papernot_limitations_2016}                                       & \begin{tabular}[l]{@{}l@{}}0.05, 0.10, 0.20, 0.25,  0.30, 0.50, 0.70, 0.90\end{tabular} \\ 
Pengcheng et al. \cite{pengcheng_query-efficient_2018} & \cite{goodfellow_explaining_2015} (FGSM)                                                                         & 0.2                                                                                       \\ 
Pal et al. \cite{pal_activethief_2020}           & \cite{goodfellow_explaining_2015} (FGSM)                                                                         & 0.25                                                                                      \\ 
Juuti et al. \cite{juuti_prada_2019}               & \cite{goodfellow_explaining_2015} (FGSM)                                                                         & 0.25 \\ 
He et al. \cite{he_drmi_2021}                   & \cite{madry_towards_2018} (PGD)                                                                                  & 0.5 \\ 

 \bottomrule
\end{tabular}
}
    \end{table}

\section{Framework for Attack Comparison}
\label{sec:framework}
In this section, we introduce our framework for attack comparison. The framework is built upon the threat model defined in the previous section. As the first step, we define how each individual attack can be evaluated in terms of effectiveness and efficiency. Then we discuss how to compare different attacks based on their threat models. 
To this end, we outline attack characteristics that should be consistent to ensure comparability and propose an attack categorisation based on them. 
Finally, we apply the proposed categorisation to prior work and use it to identify current limitations and challenges in the field.

\subsection{Attack Evaluation}
Before comparing different attacks, we first have to identify how each individual attack is evaluated. Below, we demonstrate how the attacker's goal and capabilities defined earlier in \Cref{tab:threat_model} are naturally aligned with the effectiveness and efficiency of an attack. 

\textbf{Effectiveness.}
We characterised the goal of an attack in \Cref{sec:threat_model:attack_goal} by three metrics: accuracy, fidelity, and transferability. Each of these metrics represents the effectiveness of an attack. We discussed difficulties of using transferability to compare different prior attacks in \Cref{sec:threat_model:limitations_and_challanges:goal}. Whereas it is a valid metric for evaluating the similarity of target and substitute models near their decision boundaries, it requires further agreed-upon standards to be consistently measured and comparable across different works. However, both accuracy and fidelity are unequivocally defined in the literature and can be used to compare the effectiveness of attacks. The only constraint is that the evaluation has to be performed on the same test data.

\textbf{Efficiency.} We identified in \Cref{sec:threat_model:capabilities} the capabilities of an attacker with the query budget of the attack. In \Cref{sec:threat_model:limitations_and_challanges:capabilities}, we discussed how the capabilities can be characterised by other values, for instance, the number of queries used per target model's parameter. Nevertheless, all these metrics are based on the query count and, therefore, represent the efficiency of an attack.\\ 

The attacker's goal and capabilities as defined in \Cref{sec:threat_model} explicitly correspond to the effectiveness and efficiency of an attack. Therefore, they enhance comparability by offering explicit numerical values for direct comparison. However, different assumptions about the attacker's knowledge lead to different levels of attack difficulty. 
For a fair comparison, we should first ensure that attacks are being performed under the same restrictions that originate from the attacker's knowledge. To this end, we propose a categorisation of attacks based on the attacker's knowledge, which ensures that attacks from the same category are comparable. 

\subsection{Attack Categorisation}

The second part of our framework is an attack categorisation, which ensures that attacks within the same category are comparable. 
By design, and as also suggested for other related fields, such as evasion attacks~\cite{carlini_evaluating_2019}, model stealing attacks should be examined within their threat model as defined in \Cref{sec:threat_model}. Usually, attacks need an adaptation to be launched in a broader spectrum of settings. \Cref{tab:threat_model} demonstrated how vastly threat models can differ from paper to paper. 

\begin{figure}[t]
\centering
\small
\resizebox{0.6\linewidth}{!}
{
\input{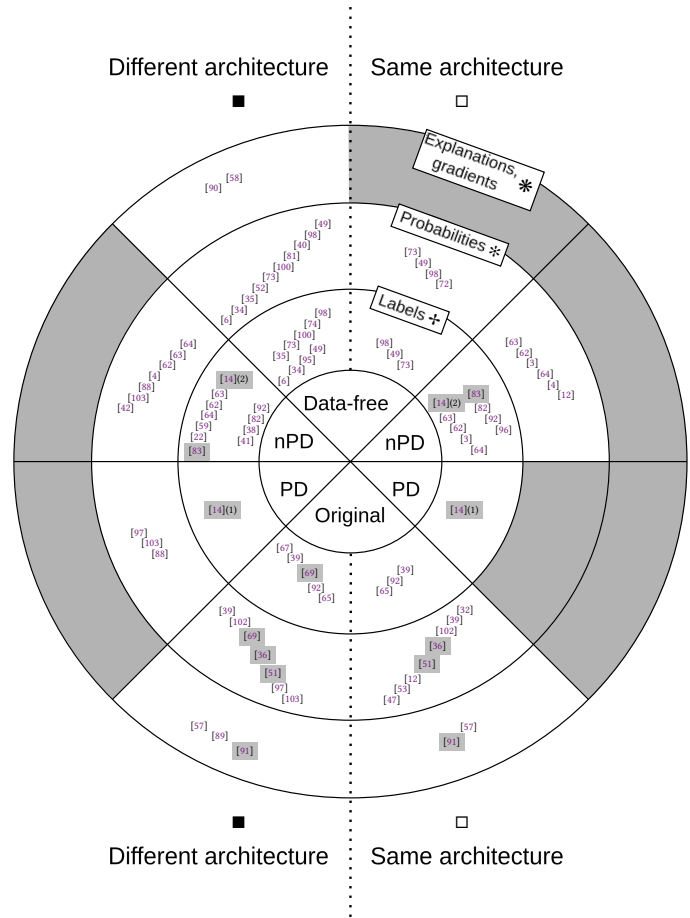}
}
\caption[Categorisation of attacks on image classifiers based on attacker's knowledge.]{Model stealing attacks against image classifiers categorised accordingly to the attacker's knowledge.} 
\label{fig:substitute_attacks_comparison}
\end{figure}

However, as was established in the previous section, the attacker's capabilities and goals are aligned with the attack evaluation. Therefore, an attack approach can be easily adapted to become comparable with others by reporting accuracy, fidelity, transferability (effectiveness), and the number of queries (efficiency). In contrast, changing the attacker's knowledge can significantly impact the methodology of an attack. For instance, providing real data to a data-free attack vanishes the primary purpose of an attack being data-free. Hence, comparing two model stealing attacks is, in principle, only meaningful if these attacks assume the same attacker's knowledge. To this end, we devised a graphical representation of the attacker's knowledge and grouped all analysed papers based on this assumed attacker's knowledge, shown in \Cref{fig:substitute_attacks_comparison}.

The diagram is constructed as follows. The dashed vertical line divides the diagram into two parts: the left half corresponds to settings where the target and substitute architectures are different, and the right half corresponds to settings that use the same architecture for both models. 
Sectors spanning the concentric circles correspond to particular knowledge about the data: the data-free assumption is on the top, followed by non-problem-domain (nPD) and problem-domain (PD) data, and original data at the bottom. 
Each sector is divided into segments by three concentric circles, which represent knowledge about the target model outputs. The inner circle corresponds to labels, the middle circle to probabilities, and the outer circle to explanations and gradients. Through this partitioning, each derived segment corresponds to a specific attacker's knowledge, covering altogether $24$ different scenarios. 

We map knowledge assumed by each paper from \Cref{tab:threat_model} into a segment corresponding to the same knowledge by placing a reference to the paper into the segment. 
If a paper covers different scenarios that fit into several segments, we put the corresponding reference in each segment. If some information about the knowledge is not given, we place its reference into each potential segment and highlight it in grey. We fill empty segments in dark grey, meaning that there are currently no works that consider the attacker's knowledge associated with those segments.

\subsection{Trends and Observations}
\Cref{fig:substitute_attacks_comparison} demonstrates the current coverage of different categories in the field of attacks against image classifiers. We identify the following trends:

(i) For a fair comparison, one should only consider attacks within the same segment, meaning that only a small fraction of works are actually comparable to each other. The largest segment (which represents different architectures, nPD data, and only labels as output) contains at most 11 papers (2 works have undefined knowledge about the target architecture), which is less than 25\% of the analysed papers.

(ii) A quarter of the segments have no prior work at all (the grey segments), a third have at most one publication, and more than half have at most three prior publications. At the same time, three segments have nearly 10 works, demonstrating a strong imbalance among threat models investigated in prior studies.

(iii) More than half of the segments with explanations as target output knowledge are not explored at all. In particular, none of the papers explore how to launch a substitute training attack having explanations (or gradients) and either PD or nPD data.

(iv) The diagram is imbalanced towards the upper left, inner part, indicating that more papers tend to assume a weaker attacker's knowledge. 
A well-performing attack with a weaker knowledge assumption raises more threats than one with a stronger assumption, as it can be successfully launched against a "more protected" API, i.e., one that does not disclose a lot of information about the target model and the original data. However, attacks with a stronger assumption are more useful for estimating the worst-case scenario when testing defences. Designing new attacks in the lower part of the diagram is, therefore, more beneficial for defence development.

(v) Complementing the previous point, we notice that the area of attacks that assume the availability of PD data is the least developed among data-related assumptions of the attacker's knowledge. However, especially for image classification, PD data can actually be the most practical scenario. 
The development of practical attacks facilitates a better understanding of which defences should be applied in real-world scenarios. Therefore, we would like to encourage the community to explore more practical scenarios.

\section{Frequent Attack Setups from Related Work}
\label{sec:attack_setup}
In the previous sections, we analysed prior work in terms of threat modelling and comparability. In this section, we study a more practical side of established attacks. We provide insightful statistical information about the examined papers to demonstrate which experimental setups were the most frequent in the recent works. In particular, we examine which image classification tasks were addressed, which target and substitute models were used, and how many queries were utilised. The overview of setups used in each work is summarised in Appendix \ref{app:setup}, whereas, in this section, we provide statistics to examine the general comparability between different works.

\subsection{Dataset}
\label{sec:attack_setup:dataset}

\begin{figure}
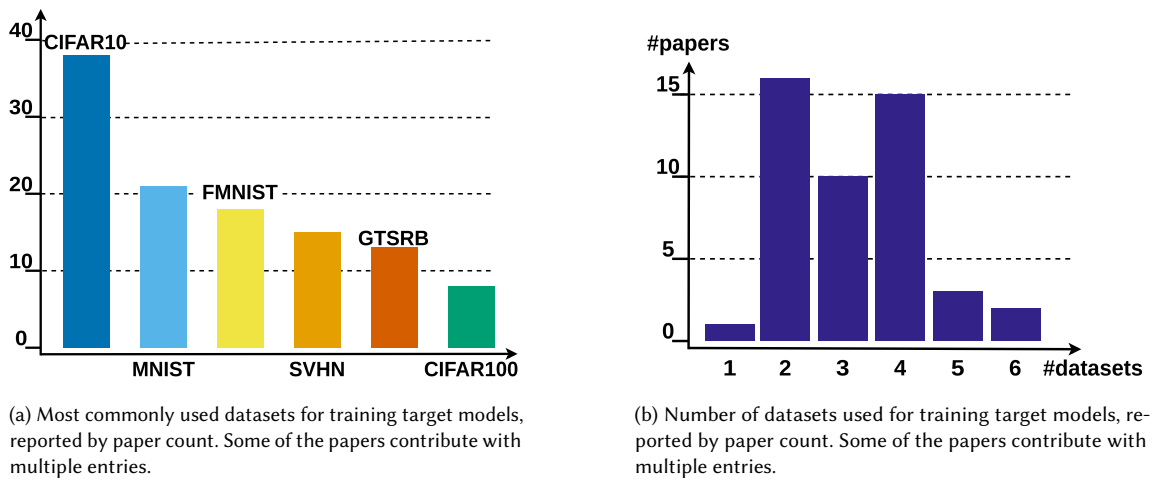

     \centering
     \begin{subfigure}{0.45\textwidth}
         \centering
         \includesvg[width=\textwidth]{images/datasets.svg}
          \caption[Datasets used for training target models.]{Most commonly used datasets for training target models, reported by paper count. Some of the papers contribute with multiple entries.} 
          \label{fig:datasets}
     \end{subfigure}
     \hfill
     \begin{subfigure}{0.45\textwidth}
         \centering
         \includesvg[width=\textwidth]{images/dataset_count.svg}
         \caption{Number of datasets used for training target models, reported by paper count. Some of the papers contribute with multiple entries.} 
         \label{fig:datasets_per_paper}
     \end{subfigure}
     \caption{Dataset statistics.}
\end{figure}

The classification task of the target model is mainly defined by its training dataset. 
We thus report the commonly used target model datasets in \Cref{fig:datasets}. If, within the same work, experiments were conducted for different original datasets, we counted the corresponding paper multiple times. 
Overall, more than 20 different datasets were used. The most popular dataset is CIFAR-10 (38 papers), followed by MNIST (21), Fashion MNIST (FMINST) (18), Street View House Numbers (SVHN) (15), German Traffic Sign Recognition Benchmark (GTSRB) (13), and CIFAR-100 (8).
Additionally, we analysed how many datasets are used for training target models in each paper, with the results presented in \Cref{fig:datasets_per_paper}. Most papers study either two datasets (16 papers) or four (15), as visible from the two peaks.

\subsection{Target Model}
\label{sec:attack_setup:target_model}
\begin{figure}
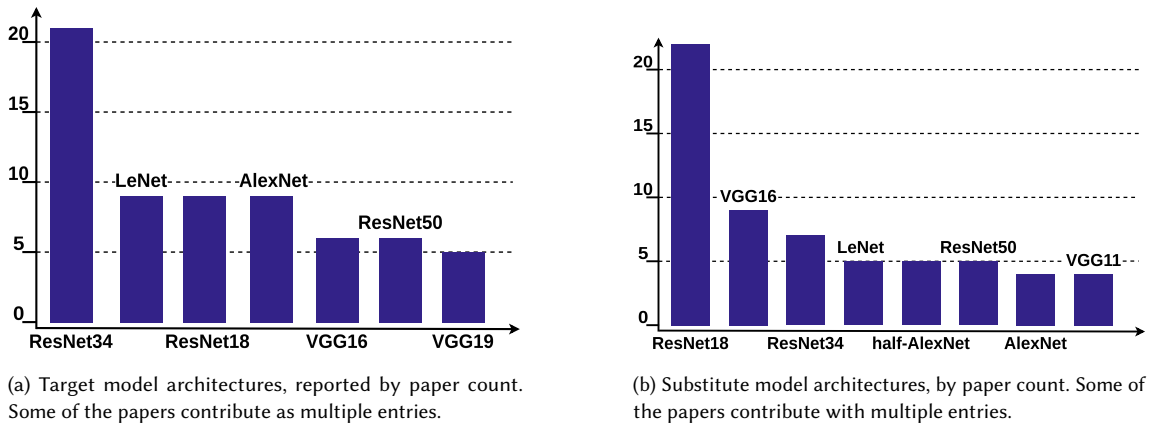

     \centering
     \begin{subfigure}{0.45\textwidth}
         \centering
         \includesvg[width=\textwidth]{images/target_models.svg}
          \caption[Target model architectures, reported by paper count.]{Target model architectures, reported by paper count. Some of the papers contribute as multiple entries.} 
         \label{fig:target_models}
     \end{subfigure}
     \hfill
     \begin{subfigure}{0.45\textwidth}
         \centering
         \includesvg[width=\textwidth]{images/substitute_models.svg}
         \caption[Substitute model architectures, reported by paper count.]{Substitute model architectures, by paper count. Some of the papers contribute with multiple entries.} 
         \label{fig:substitute_models}
     \end{subfigure}
     \caption{Target and substitute model statistics.}
\end{figure}

Another important factor besides the dataset is the target model architecture; statistics of usage of popular architectures are shown in \Cref{fig:target_models}. As with datasets, if several target models were trained in one paper, they were counted multiple times. The most popular choices are ResNet34 (20), AlexNet (9), LeNet (9), ResNet18 (9), VGG16 (6), ResNet50 (6), and VGG19 (5).
However, the information presented is not complete: in six papers, information about the target model was (at least partially) missing, and in 13 papers, custom CNN architectures were utilised. 

Subsequently, we aligned the most popular dataset choices with the target architectures to examine the relationship between those two choices. \Cref{tab:datasets_vs_target} demonstrates how many times the most common target architectures were used to learn the most common datasets. 
Each paper can contribute multiple times, as we counted the total number of model-dataset combinations across all 47 works. As a result, some papers contributed significantly more entries than others. For example, Yan et al.~\cite{yan_holistic_2023} trained 16 different target models for each of four target datasets (CIFAR-10, MNIST, FMNIST, CIFAR-100), contributing 64 entries to \Cref{tab:datasets_vs_target}. For brevity, we only list the top 7 most common architectures. Yan et al.~\cite{yan_holistic_2023} experimented with 5 of them, meaning that the remaining 11 architectures contributed to the last row of each of the four datasets in \Cref{tab:datasets_vs_target}.

Only for the SVHN dataset, the majority of experiments were conducted using one of the listed architectures, namely ResNet34. However, if we deduct the 11 entries of Yan et al.~\cite{yan_holistic_2023} from the last row (who use CIFAR-10, MNIST, FMNIST, and CIFAR-100), the ResNet34 model is also the most common option for CIFAR-10 and CIFAR-100. For MNIST and FMNIST, among the listed architectures, LeNet is the most typically used. For GTSRB, the most frequently used architecture is a custom CNN (see Appendix \ref{app:setup} for more details).

\begin{table}[ht!]
\centering
\caption{Frequency of usage of a specific architecture for training target models on various datasets. Some of the papers contribute with multiple entries.}
\label{tab:datasets_vs_target}
\begin{tabular}{lccccccc}
\toprule
                & CIFAR-10 & MNIST & FMNIST & SVHN & GTSRB & CIFAR-100 & Others \\ 
                \midrule
ResNet34 & 17               & 2              & 2               & \textbf{11}   & 1              & 4                 & \textbf{16}     \\ 
ResNet18 & 6                & 1              & 3               & 1             &                & 3                 & 4               \\ 
LeNet    &                  & 5              & 6               &               &                &                   & 1               \\ 
AlexNet  & 5                &                & 3               &               &                &                   & 5               \\ 
VGG16    & 3                & 2              & 3               &               & 1              & 1                 & 2               \\ 
VGG19    & 1                & 3              & 4               &               &                & 1                 & 3               \\ 
ResNet50 & 4                & 1              & 1               &               &                & 1                 & 1               \\ 
Others   & \textbf{24}      & \textbf{25}    & \textbf{22}     & 4             & \textbf{7}     & \textbf{13}       & 12              \\ 
\bottomrule
\end{tabular}

\end{table}

Finally, we examined the adoption of transfer learning for training target models. Only 7 out of 47 examined papers explicitly mention whether the target model was trained from scratch or from pre-trained weights. Only a single work reported training the target model from scratch, three reported using pre-trained weights, and three more works adopted both strategies. 

\subsection{Substitute Model}
\label{sec:attack_setup:substitute_model}
Subsequently, we investigated substitute model architectures with the results presented in \Cref{fig:substitute_models}. Here, ResNet18 (22) is the most widely used architecture,
followed by VGG16 (9), ResNet34 (7), LeNet (5), half-AlexNet (5), which is a customised version of AlexNet with half the capacity, ResNet50 (5), AlexNet (4), and VGG11 (4). As with the target model, 13 papers used custom CNN architectures. Four papers did not report the substitute architecture for at least some datasets. 
As with target architectures, we examined the usage of transfer learning for training substitute models. Whereas most works (29) do not explicitly mention from which weights the substitute model is trained, the number of works that report this information (18) is significantly larger than for target models (7). Among those 18 works, 9 train substitute models from scratch, 7 start with pre-trained weights, and two works investigate both approaches. 

We present how often the most popular substitute architectures were used to steal the most popular target architectures in \Cref{tab:target_vs_substitute}. As stated earlier, the contribution of different papers varies significantly from just a single entry to 64 entries (four substitute models against four target models trained for each of four datasets~\cite{yan_holistic_2023}). The latter has significantly contributed to the "Others/Others" setting (right bottom cell), as a significant part of the combinations contains architectures that are not that common. However, unlike with the datasets, for most of the target architectures, there is a match from the substitute list. ResNet18 was mostly used for stealing ResNet34 and ResNet18, LeNet and VGG16 were used to steal themselves, and for AlexNet the most common substitute choice was half-AlexNet. For VGG19 and ResNet50, 
 \Cref{tab:target_vs_substitute} is inconclusive.

\begin{table}[ht!]
\centering
\caption{Frequency of usage of a specific architecture for training substitute models (rows) to attack various target models (columns). Some of the papers contribute with multiple entries.}
\label{tab:target_vs_substitute}
\begin{tabular}{lcccccccc}
\toprule
 & ResNet34    & ResNet18    & LeNet      & AlexNet    & VGG16      & VGG19      & ResNet50    & Others      \\ \midrule
ResNet34                                                                                    & 9           & 1           &            & \textbf{}  & 1          &            & 1           & 3           \\ 
ResNet18                                                                                    & \textbf{16} & \textbf{10} &            & 3          & 1          &            & 3           & 6           \\ 
LeNet                                                                                       &             &             & \textbf{5} & 2          &            &            &             & 2           \\ 
AlexNet                                                                                     & 1           & 2           & 2          & 4          & 1          &            &             & 3           \\ 
VGG16                                                                                       & 4           &             & 1          &            & \textbf{5} &            & 1           & 4           \\ 
VGG11                                                                                       & 2           & 2           &            & 2          & 2          &            & 1           & 6           \\ 
RN50                                                                                    & 2           &             &            &            & 1          &            & 1           & 7           \\ 
half-AlexNet                                                                                &             &             &            & \textbf{7} &            &            &             &             \\ 
Others                                                                                      & 4           & \textbf{}   & \textbf{5} &            & 2          & \textbf{3} & \textbf{12} & \textbf{97} \\ \bottomrule
\end{tabular}

\end{table}

\subsection{Number of Queries}
\label{sec:attack_setup:number_of_queries}
\begin{figure}[t]
\centering
\includesvg[width=\linewidth]{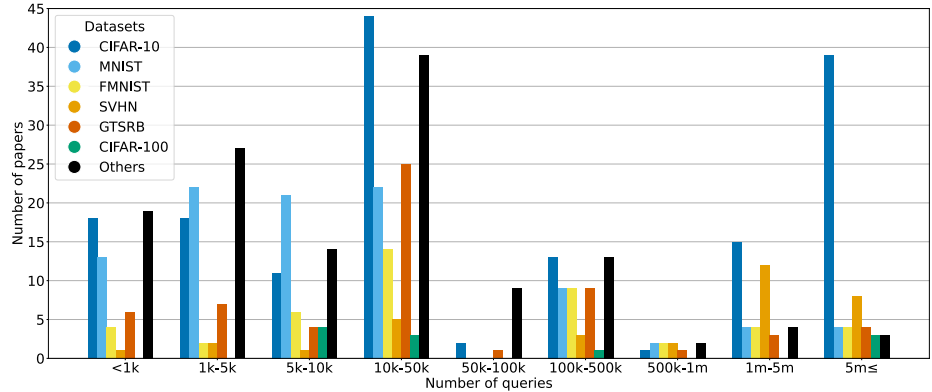}
\caption{Number of queries reported in papers targeting models trained on various datasets. Some of the papers contribute with multiple entries.} 
\label{fig:queries_all}
\end{figure}

The last attack setup component we analysed is the number of queries reported. We plot the frequency of using a certain number of queries in \Cref{fig:queries_all}, highlighting the datasets from \Cref{fig:datasets}. The number of queries is binned into different groups for better visual comprehension. Most of the experiments were conducted with fewer than 50,000 queries, with especially many in the range between 10,000 and 50,000 queries. Barely any results were reported for two groups: between 50,000 and 100,000 queries and between 500,000 and 1,000,000 queries. For CIFAR-10 and SVHN, a lot of experiments were performed with over 1 million queries. Further investigation revealed that this trend is caused by data-free attacks. Unlike attacks that use real images to train a substitute model, attacks that rely on artificial data usually require millions of queries to converge~\cite{hong_exploring_2023, yang_efficient_2023}. In turn, attacks that utilise real images are constrained by the size of the dataset from which the images come. 

\begin{figure}[t]
\centering
\includesvg[width=\linewidth]{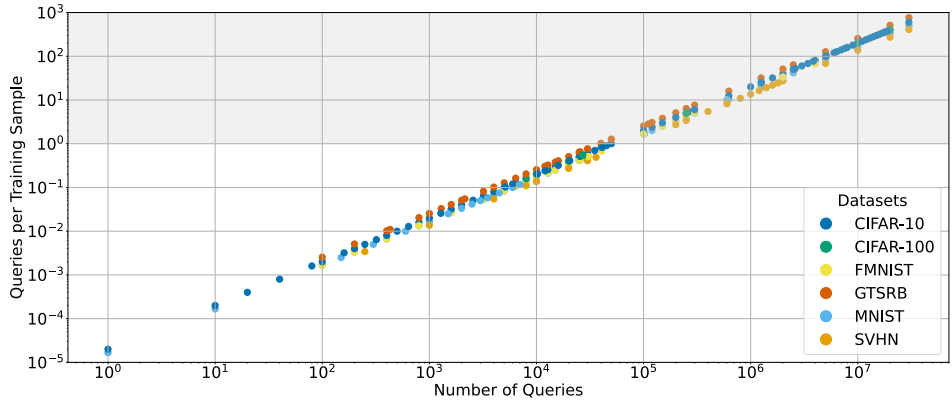}
\caption{Queries per target model's training data sample, plotted against the total number of queries. The grey area highlights attacks that use more than one query per target model's training sample. Some of the papers contribute with multiple entries.} 
\label{fig:queries_relative}
\end{figure}

While \Cref{fig:queries_all} gives an overall picture, the number of queries is not representative enough for comparability purposes (see our discussion in \Cref{sec:threat_model:limitations_and_challanges:capabilities}). For this reason, we calculated the ratio between the number of queries and the target model's training set size. This ratio shows how many queries an attacker spends per sample from the target model's training data. We plot it against the number of queries in \Cref{fig:queries_relative}. 

Only for two datasets, MNIST and CIFAR-10, was a very low number of queries (below 100) studied. Most of the works, as stated above, used up to 50,000 queries, with 0.01 to 1 queries spent per target model's training data sample. Those attacks, therefore, used at most as many samples as there are in the target model's training dataset, making them relatively efficient. On the other side, data-free attacks with millions of queries spent correspond to at least 100 queries per training sample. 

\section{Best Practices}
\label{sec:best_practices}
In this section, we summarise our observations from prior sections into guidelines for designing and evaluating model stealing attacks. We gather best practices from previous work that can be performed before, after, and beyond experiments and present them as a list of recommendations (Rs). 

\subsection{(R1) Before Experiments}
\label{sec:best_practices:before_exp}
We begin by discussing recommended strategies for developing and evaluating model stealing attacks before the experimental part starts.
In particular, we emphasise the importance of a clear threat model and an adequate experimental setup. 

\textbf{(R1.1) Threat model.} 
Well-defined threat models are essential for ensuring attack comparability and countermeasure development. In this work, we primarily focused on attacks that intend to copy the behaviour of the target model, for which we defined a detailed overview of the attacker's profile in \Cref{sec:threat_model}. For other attackers' goals, we direct readers to a related survey~\cite{oliynyk_i_2023}. 

For behaviour-stealing attacks based on substitute model training, the following aspects of the attacker's knowledge and capabilities must be clearly specified:
\begin{itemize}
    \item \textit{Attacker's data.} As we defined in \Cref{sec:threat_model:knowledge}, an attack can use original, problem-domain, non-problem-domain data or no real data at all (i.e., data-free attack). 
    \item \textit{Target model's architecture.} Knowing the target model's architecture can significantly simplify the selection of the substitute model's architecture: using exactly the same architecture for both models leads to higher transferability scores~\cite{juuti_prada_2019}, giving even more advantage to the adversary.
    \item \textit{Pre-trained models.} As discussed in \Cref{sec:open_research_questions:models}, pre-trained models can give a significant performance improvement to the adversary. Therefore, authors should specify whether the target, substitute, or any auxiliary model is pre-trained and should report which dataset was used.
    \item \textit{Target model's outputs.} For classification problems, we differentiate between labels, probabilities (confidence score), and gradients or explanations as outputs, as was defined in \Cref{sec:threat_model:knowledge}. Prior works have demonstrated that having outputs that reveal more information leads to better-performing attacks: probabilities give better results than labels~\cite{orekondy_knockoff_2019, juuti_prada_2019, atli_extraction_2020}, and explanations render better models than probabilities~\cite{miura_megex_2021}.
    \item \textit{Query budget.} All limitations and capabilities of the attacker in terms of query budget numbers must be explicitly stated. In addition, if an attack requires a validation set labelled by the target model, its size should ideally be included in the total query budget, as done by Jindal et al.~\cite{jindal_army_2024}.
\end{itemize}

\textbf{(R1.2) Experiment Setup.}
Defining an appropriate experimental setup is another crucial step before launching the experiments. In particular, it should correspond to the defined attacker's knowledge and capabilities and be sufficient to benchmark the results against the state-of-the-art. The statistics reported in \Cref{sec:attack_setup} can serve as a good starting point for defining attack setups. 
\begin{itemize}
    \item \textit{Target dataset.} As we mentioned earlier in \Cref{sec:open_research_questions:datasets}, large and complex datasets have not been widely studied before. Therefore, researchers should explore a variety of target datasets with different complexities. Besides, if the baselines defined in the previous step are earlier attacks, using the same target datasets is crucial for comparability. 
    \item \textit{Attacker's data.}
    The attacker's dataset must align with the defined threat model. In \Cref{sec:open_research_questions:datasets}, we discussed that semantically, the same attacker's knowledge about data could be interpreted differently. Therefore, following previous work~\cite{orekondy_knockoff_2019}, researchers must verify whether the attacker's data has any (unintentional) intersections with the original target model's training (or test) data. Furthermore, removing common data from the attacker's set can potentially make the actual setup closer to the defined threat model. Additionally, authors must specify whether any additional data was used for training an auxiliary model~\cite{lin_quda_2023, yang_efficient_2023}.
    \item \textit{Target model.} 
    Similarly to datasets, complex state-of-the-art models are under-represented in the current research, as was mentioned in \Cref{sec:open_research_questions:models}. While exploring more complex models is beneficial for evaluating attack generalisability, simpler models might be needed for comparability with previous work (see \Cref{sec:attack_setup:target_model}). For a fair comparison, either exactly the same target model can be used~\cite{rosenthal_disguide_2023}, or the target model can be trained until it achieves approximately the same performance as in previous work~\cite{sanyal_towards_2022}.
    \item \textit{Substitute model.} 
    The choice of substitute model should align with the assumed knowledge about the target model, i.e., if the target architecture is unknown, the substitute architecture should be different. 
\end{itemize}

\subsection{(R2) During Experiments}
\label{sec:best_practices:during_exp}
We continue the discussion with best practices that should be applied during model stealing experiments. At this stage, a proper evaluation of attacks is the most crucial. We approach it from several perspectives: defining appropriate baselines, performing exhaustive effectiveness and efficiency evaluation, and conducting ablation studies. 

\textbf{(R2.1) Baselines.}
Defining appropriate baselines is essential for aligning novel attacks with the state-of-the-art. We recommend using \Cref{tab:threat_model} and \Cref{fig:substitute_attacks_comparison} to identify state-of-the-art attacks with an identical threat model. Among others, these attacks are the most feasible and fair baselines for a novel attack. 

\textbf{(R2.2) Effectiveness evaluation.}
As was demonstrated in \Cref{tab:threat_model}, none of the metrics (accuracy, fidelity, transferability) was reported for every previous attack. While accuracy is the most common, it is only meaningful when the accuracy of the target model is specified as well. Fidelity is more self-explanatory, but it is often missing from the attack assessment (\Cref{tab:threat_model}). Therefore, to ensure a comprehensive effectiveness evaluation, we recommend including both metrics.

As discussed in \Cref{sec:threat_model:limitations_and_challanges:goal} and illustrated in \Cref{tab:transferability}, there is no common practice on how transferability should be measured for model stealing attacks. Previous work disagrees on both methods for crafting adversarial examples and samples on which adversarial perturbations are applied.  
However, transferability is heavily investigated in the adversarial example community~\cite{papernot_transferability_2016,gu_survey_2023}. In contrast to model stealing, research has brought forward benchmarks~\cite{croce_robustbench_2021} and related strong attacks~\cite{croce_reliable_2020} as well as research on failure modes of attacks~\cite{pintor_indicators_2022}. Evaluations for model stealing should draw from this literature to obtain more reliable evaluations. 

Finally, if the effectiveness evaluation is performed on a dataset with a predefined test set, the latter should remain unmodified. While the test set could be a good source of original data used for an attack, leaving only a fraction of it for actual testing may lead to some discrepancies. In general, this may not invalidate an attack, but it makes the comparison with previous attacks difficult, especially if the difference in performance is not significant. 

\textbf{(R2.3) Efficiency evaluation.}
The primary indicator of the attack efficiency is the number of queries used to train a substitute model. Several sources can be used to identify potentially relevant values: (i) the pre-defined query budget (as discussed in  \Cref{sec:open_research_questions:queries}), (ii) convergence of the substitute model~\cite{hong_exploring_2023}, (iii) number of queries reported by baseline state-of-the-art attacks~\cite{truong_data-free_2021}, and (iv) number of queries required to achieve the same performance as the baselines~\cite{rosenthal_disguide_2023}. 
However, as mentioned earlier in \Cref{sec:open_research_questions:queries}, placing the number of queries in context with the complexity of the target model~\cite{oliynyk_i_2023} or its learning task (as was done in \Cref{sec:attack_setup:number_of_queries}) gives a more objective efficiency assessment.

\textbf{(R2.4) Ablation studies.} 
Ablation studies are essential for understanding the contribution of each attack component and verifying whether attacks that incorporate additional techniques for performance improvement indeed perform better. From a certain perspective, they also produce baselines in the form of a basic attack on top of which the improvements were built. Therefore, authors should perform an ablation study for every (new) attack parameter.  Good examples include (i) comparing an attack that uses two data types with an attack that only uses one type of data~\cite{correia-silva_copycat_2018}, (ii) evaluating how query optimisation techniques improve the attack performance compared to the non-optimised attack~\cite{jagielski_high_2020, pal_activethief_2020}, and (iii) comparing the performance of an ensemble of substitute models with a single substitute model~\cite{jindal_army_2024}.

\subsection{(R3) Beyond Experiments}
\label{sec:best_practices:beyond_exp}
Finally, we outline additional best practices that go beyond standard experiments. They include validating attacks in practical scenarios, testing their ability to evade state-of-the-art defences, reporting negative results, ensuring the reproducibility, and disclosing vulnerabilities in an ethical way.  

\textbf{(R3.1) Real-world APIs.}
Real-world applications are commonly not a feasible setting for an adequate attack evaluation. 
Stealing a model, even for scientific purposes, may destroy the intellectual property of the owner of this model~\cite{foss-solbrekk_three_2021}, raising ethical issues.
In addition, APIs usually operate on a black-box basis, hiding information about the (dynamically changing) target model and its learning task. 
Furthermore, deployed models often have constraints on the allowed queries~\cite{grosse_towards_2024}, potentially rendering attacks with huge query budgets infeasible. These restrictions can narrow down or even modify the initial threat model of an attack, leading to a less comprehensive evaluation. Moreover, some metrics might be impossible to measure, for instance, the accuracy of the substitute model on the original learning task or the efficiency score (the number of queries divided by the number of weights in a model). However, these conditions correspond to a real attack scenario, and understanding how dangerous attacks are against real-world APIs and to what extent they can actually be measured is vital for the development of countermeasures. Therefore, following previous works~\cite{correia-silva_copycat_2018, yu_cloudleak_2020, gong_inversenet_2021}, we highly recommend evaluating novel approaches against real-world applications, while ensuring solid, i.e., non-evolving, setups that ensure interoperability.

\textbf{(R3.2) Defences.}
Another important step in further attack evaluation is testing them against the state-of-the-art defence mechanisms. To ensure fair evaluation, it is crucial to select defences specifically designed for a concrete threat model. We refer readers to the recent survey for further information about defence categories and the line-up between attacks and defences~\cite{oliynyk_i_2023}. In addition, general guidelines on how to evaluate defences should be developed, analogously to evasion attacks~\cite{carlini_evaluating_2019}, where testing against adaptive attacks is a de facto  standard~\cite{carlini_evaluating_2019}.

\textbf{(R3.3) Negative results.}
Additionally, we would like to encourage the community to share negative results. While often not as groundbreaking as positive results, they might provide key insights for future work. This is especially important for results related to countermeasure testing, as robust defence mechanisms are the common ultimate goal of the community. Furthermore, reporting negative results safeguards others from re-investing in the same research, saving computational and time costs. In turn, publications of negative results should be encouraged by the program chairs of publication venues. 

\textbf{(R3.4) Reproducibility.}
Reproducibility ensures comparability and enables future follow-up work. To this end, design and implementation processes need to be reported in detail. In particular, all components mentioned above, such as threat model, experiment setup, and evaluation, must be provided, either with the code or in the research output. The implementation should be publicly available with explicit documentation and instructions on how to recreate the experiment results.

\textbf{(R3.5) Ethics.} At the same time, releasing code may make it easier for malicious entities to perform an attack. Along these lines, it is of general importance to comply with best practices of disclosure\footnote{\url{https://cheatsheetseries.owasp.org/cheatsheets/Vulnerability_Disclosure_Cheat_Sheet.html}} when a model is found to be particularly vulnerable to be stolen. Calls to implement the corresponding infrastructure for both vulnerabilities~\cite{longpre_position_2024} and incidents~\cite{bieringer_position_2025} have been made in the scientific community.

\section{Open Research Questions}
\label{sec:open_research_questions}
In this section, we discuss the open research questions (RQs) that emerge from our analysis. We organise them by first reviewing open questions relating to datasets, models, and then the number of queries.

\subsection{(RQ1) Datasets}
\label{sec:open_research_questions:datasets}
We identify several open research questions associated with the datasets used. First, we discuss open questions related to the used datasets and attack generalisation across datasets. Afterwards, we demonstrate the current lack of a formal definition of attacker's knowledge about data. 

\textbf{(RQ1.1) Generalisation across datasets.} We start with a discussion about the used datasets and attack generalisation across these.
The statistics presented in \Cref{sec:attack_setup:dataset} show that datasets like CIFAR-10, MNIST, and Fashion MNIST are most commonly studied. Yet, attacks launched against MNIST classifiers perform in general better than against more complex datasets, for instance, GTSRB~\cite{juuti_prada_2019, pal_activethief_2020}, SVHN~\cite{yuan_es_2022, khaled_careful_2022} or CIFAR-10~\cite{yang_efficient_2023}. A noise-based attack by Milli et al.~\cite{milli_model_2019} only works on MNIST classifiers, and is ineffective against CIFAR-10 classifiers, suggesting that attacks do not necessarily generalise.
These findings underline the need to study more challenging tasks, such as CIFAR-100, which is currently only studied in 8 papers, being the 6th most frequent target dataset, compared to 45 (out of 47 examined) papers studying (Fashion) MNIST or CIFAR-10.

\textbf{(RQ1.2) Attacker's knowledge about data.} Further open questions related to dataset complexity occur due to vaguely defined attackers' knowledge about the training data.  
For example, Truong et al.~\cite{truong_data-free_2021} demonstrated that the similarity of attacker's and original data domains significantly contributes to the attack effectiveness. On the other hand attacker data types are not clearly defined as we derived in \Cref{sec:threat_model:limitations_and_challanges:knowledge}: (i) current works have different understandings on what the original data consist of (the exact training data of the target model or the distribution this data was drawn from), (ii) data that at first sight looks like non-problem domain data can turn out to be problem-domain due to unintentional intersections between data domains, and (iii) in literature, a data-free attack stands for both an attack that does not use any real data to query target model and an attack that does not use any real data to train a substitute model, whereas querying with real data is possible to e.g. train an auxiliary generative model. In all these examples, semantically the same knowledge about the data can give more or less benefits for an attack. For a fair comparison, this should not be the case. Consequently, more research and, ultimately, standards are required to better distinguish differences in attacker's knowledge types and understand their impact on attack success. 

\textbf{(RQ1.3) Overlap across datasets.} Finally, it remains also an open question to which degree data is shared or accessible in practice~\cite{grosse_machine_2023}. We have loosely discussed this point when discussing the data originally used to train the models (cf. \Cref{sec:attack_setup:target_model,sec:attack_setup:substitute_model}). In real world applications, it remains an open question which data preprocessing steps have been carried~\cite{grosse_towards_2024,grosse_machine_2023} out and how they affect potential dataset overlap when the same data is used but slightly changed due to pre-processing. An addition open question remains whether data was sourced from public sources and is thus accessible indirectly to the attacker~\cite{grosse_towards_2024}. Overlap in the data used should thus be object of study and documented in detail.

\subsection{(RQ2) Models}
\label{sec:open_research_questions:models}
We further present open questions related to various models that play a role in model stealing attacks, namely target models, substitute models, and auxiliary models such as generators. We start with discussing current limitations in attacks against complex, state-of-the-art models. Then, we outline open questions regarding the usage of pre-trained substitute and auxiliary models. Finally, we review partial stealing of models and whether it is a simpler and, therefore, more dangerous task.  

\textbf{(RQ2.1) Large and complex models.} We begin with a discussion about attacks against complex state-of-the-art models. In \Cref{sec:attack_setup:target_model}, we presented the most common target architectures (with a comprehensive overview in Appendix \ref{app:setup}). However, none of these works studies more advanced architectures such as EfficientNet~\cite{tan_efficientnet_2019}, BigTransfer (BiT)~\cite{kolesnikov_big_2020}, or vision transformers (ViT)~\cite{dosovitskiy_image_2021}, which currently lead in CIFAR-10 classification task\footnote{\url{https://paperswithcode.com/sota/image-classification-on-cifar-10}}. These architectures have not been studied neither in the context of other model stealing attacks, nor for training a substitute model~\cite{oliynyk_i_2023}. Further research is needed to determine whether current attacks scale to more advanced and complex architectures. 

This is especially relevant as recent work suggests that attacks might not easily generalise to complex models. Gudibande et al.~\cite{gudibande_false_2024} demonstrated that imitation of ChatGPT through fine-tuning open-source large language models is not successful. Although fine-tuned models learned to mimic the style of ChatGPT, with further evaluation, it turned out that there is a significant gap in the factuality of ChatGPT and fine-tuned models.

\textbf{(RQ2.2) Pre-trained models.} Another key question concerns the usage of pre-trained models by adversaries.
Zhang et al.~\cite{zhang_thief_2021} demonstrated that the availability of pre-trained substitute models could give a significant advantage to an adversary compared to models trained from scratch. Such an overlap is for example created when the target and substitute models were both pre-trained on the same data, which is also a subset of the training data of the target model. This concern has practical relevance, as there is evidence that pre-trained models are heavily used when deploying AI in industry~\cite{grosse_towards_2024}. However, as we mentioned earlier in \Cref{sec:attack_setup:target_model} and \Cref{sec:attack_setup:substitute_model}, information on whether pre-trained weights were used for target or substitute models is often omitted. 

Pre-trained models can also be implicitly used in the attack, for instance, to initialise a synthetic data generator~\cite{barbalau_black-box_2020, yang_efficient_2023, lin_quda_2023}. In this case, an attack implicitly relies on the knowledge that the generator absorbed from its pre-training data. Even if only generated synthetic data is further used to train a substitute model, there is a significant hidden contribution of that preliminary generator's knowledge. Therefore, the usage of pre-trained models needs to be explicitly communicated and potentially included in the attacker's knowledge and capabilities.

\textbf{(RQ2.3) Partial model stealing.} Finally, we turn our attention to the problem of partial stealing. In real-world scenarios, an adversary might be interested only in a specific behaviour of the target model. In this case, the threat model defined in \Cref{sec:threat_model} needs to be modified, as (i) not all samples from the original data distribution are relevant, (ii) the substitute architecture needs to be modified accordingly to the goal task, and (iii) evaluation should be performed on task-relevant data.  While very limited research has been done in this direction, current findings suggest that partial stealing can be easier and, therefore, potentially more dangerous than obtaining the full functionality. Okada et al.~\cite{okada_special_2020} considered a special scenario of model stealing attacks against image classifiers, in which a substitute model is trained to replicate the target's behaviour only on a specific subset of classes. The results demonstrated that within the same query budget, the substitute model achieves higher scores when duplicating only partial functionality compared to full functionality~\cite{okada_special_2020}. This finding suggests that partial stealing is effective while being more efficient than full functionality stealing. It remains an open question if partial stealing is always a threat and if it can be countered with the same methods that are applicable against complete functionality stealing.

\subsection{(RQ3) Queries}
\label{sec:open_research_questions:queries}
The last group of open questions concern the attacker's queries.
We begin with a discussion around two similar terms: query budget and number of queries, which we will define below. Next, we analyse how practical the exact values from related work are and raise the question of their scalability. Finally, we outline the limitations of counting queries for attack efficiency assessment.

\textbf{(RQ3.1) Number of queries vs query budget.} As a starting point, we pose a question regarding understanding what query budget and number of queries characterise. While sometimes these terms are used interchangeably, we see a clear distinction between them. The number of queries is measured by an adversary as the query count required to perform their attack; it is defined by internal attack-specific constraints, e.g. how many queries are needed for the substitute model to converge.   
The query budget is, in contrast, a rather external constraint of how many queries an adversary can potentially make. There can be different sources of such constraints: limitations of the API, limited financial abilities of an adversary, monitoring defences~\cite{juuti_prada_2019} that will cut off the adversary after a certain time, etc. 
To take full advantage of an attack, the number of queries it requires should not exceed the query budget. 
However, as there is no clarity in the literature concerning this terminology, different works use different sources of constraints (external vs internal) to define how many queries to spend on their attack. Driven by different incentives, those attacks essentially correspond to different threat models. The comparability of such attacks remains an open question, as well as a clear distinction of what impacts the decision to spend that exact amount of queries. 

\textbf{(RQ3.2) Practical query numbers.} The next open question concerns the practicality of a number of queries considered in the literature. For that, we need practical query budgets, i.e. external constraints on the query count. 
Query numbers on deployed AI models, according to a recent survey, could be as low as less than 100 or less than 1,000; where in most cases, models cannot be queried at all~\cite{grosse_towards_2024}. While the authors~\cite{grosse_towards_2024} do report unlimited amounts of queries as the second most frequent, these numbers may be reduced by applying a monitoring defence that can cut off suspicious clients after as little as 100 queries~\cite{juuti_prada_2019}. 

Despite such strict practical constraints, the numbers required by attacks can be significantly higher. As shown in \Cref{fig:queries_all}, only a small fraction of works relied on less than 1,000 queries. Most of the experiments were conducted with budgets between 10,000 and 50,000 queries. Finally, data-free attacks, as noticed in \Cref{sec:attack_setup:number_of_queries}, usually require millions of queries, which likely renders them useless under practical limitations. Therefore, more studies with very limited attacker's capabilities are needed to understand how dangerous model stealing attacks are in practice. On the other side, we need to expand our knowledge about the practical scenarios to better identify potential threats.

\textbf{(RQ3.3) Query scalability.} Another practical issue with model stealing attacks is their scalability. 
In \Cref{sec:attack_setup:number_of_queries}, we demonstrated that, on average, data-free attacks need more than 100 queries per target model's training sample to steal a model. We can rephrase this observation in the following way: on average, it takes at least 100 queries to compensate for the knowledge that the target model obtains from a single training sample. The training data, in this case, consists of low-resolution CIFAR-10 images. It is an open question if such data-free attacks would scale for more complex, high-resolution data. At the moment, especially taking into account the practical number of queries reported above, such data-free attacks do not appear to be a practical threat against complex tasks.

\textbf{(RQ3.4) Other efficiency metrics.} Lastly, we would like to raise the question of whether measuring the number of queries is sufficient for efficiency evaluation. One key factor that can be taken into account is the complexity of the target model. Previous work~\cite{oliynyk_i_2023} suggested calculating a so-called efficiency score that shows how many queries per learnable parameter of the target model an attack requires. Another factor that we considered in this work in \Cref{sec:attack_setup:number_of_queries} is the complexity of the target model's task, which we quantified as the number of samples in the training set. Including both model and data complexity in efficiency assessment can be a further step here. 

Another efficiency estimation can come from the actual cost spent to train a substitute model. Previous work that, in particular, studied attacks against public APIs reported the price they had to pay for the executed queries~\cite{yu_cloudleak_2020}. While this price gives a good estimate of how efficient the attacker's data-gathering process is, the costs of actually training a model and finding appropriate architecture and hyperparameters are not included. To the best of our knowledge, estimating the actual price of creating a model from scratch is still an open question. This leads to difficulties in estimating the actual attack cost and whether it is reasonable compared to the price of creating a target model.

\section{Discussion on Generalisation}
\label{sec:discussion}
In this work, we analysed the largest and most developed group of attacks that clone the behaviour of target models offered as a service - attacks on image classification models. In the following, we discuss how our analysis, framework and recommendations transfer to other problem domains. 

\textbf{Threat model.} In \Cref{sec:threat_model} we defined the threat model for attacks against image classification models in terms of attacker's knowledge, capabilities, and goals. In the following, we discuss how each of this components can be adapted to other problem domains. The attacker's knowledge consists of four components: target model's data, its outputs, architecture, and availability of pre-trained models. Attacker's data defines potential queries and is required for any other domain as well. Similarly to image classification, one can further define different strengths of attacker's knowledge about the data depending on the similarity of available data to the target model's original training data. However, different data types may affect the attacker by adding or removing constraints~\cite{grosse2024qualitative}. As an example, tasks like trajectory prediction, representing the trajectory of a physical vehicle, are more constrained regarding attack perturbations than images with real-valued pixels~\cite{grosse2024qualitative}. An exploration of such constraints is, due to the lack of corresponding approaches, left for future work.

Regarding the knowledge of the model's output, the target model returns outputs regardless of the domain; the granularity of the outputs is however defined by the domain problem and might differ from the one for image classification. Extending, merging, or adapting the three classes provided in our work may thus be necessary, as, for example, pixel-wise tasks like segmentation by default leak more information than classification. 
Further, the notion of attacker's knowledge about the target and pre-trained models transfers across different domains in a straightforward way. Attacker's capabilities, defined by the number of queries, also transfer to any query-based attack regardless of the problem domain: while the scale of queries can vary, the concept remains the same. 

Finally, attacker's goal in this work was defined by three metrics: accuracy, fidelity, and transferability. We can view these metrics as (i) one measuring the performance of the substitute model on the task the target model was trained for, (ii) one measuring the similarity of performances of the target and substitute models, and (iii) one measuring the susceptibility of the target model to the adversarial examples crafted to fool the substitute model -- or in other words, the similarity between the two models near their decision boundaries. 
However, the measurement of performance varies largely across tasks. For example, while in vision, accuracy is defined on a set of images, in trajectory prediction, it rather considers meters of deviation over a set of trajectories and their ground truth. This definition of success affects the attacker's goal and how this goal is assessed, as in other attacks on AI~\cite{grosse2024qualitative}. The framework may thus have to be adapted in terms of measurements of success as well.
In general, however, the defined threat model can be adapted for other problem domains. Subsequently, our framework for attack comparison that follows from the threat model can also be transferred.

\textbf{Best practices and open research questions.} All of the recommendation (R1-R3) from best practices presented in \Cref{sec:best_practices} are generic and non-domain-specific and, therefore, should also be taken into account for attacks in other problem domains. The open research questions raised in \Cref{sec:open_research_questions} are derived from analysis of the particular group of attacks, and hence are more domain-specific. While some of these issues generalise across domains, and attacks in other problem domains are  less established and systematised, the relevance of these questions for other domains should be studied separately. 

\section{Related Work}
\label{sec:related_work}
Below, we review influential works on evaluation methodology with similar contributions to other research fields.

Evaluation methodology is an important aspect in any discipline utilising experimentation, and as a field matures, research methodology becomes more precise with standards and best practices emerging. In security, the basis of any methodology is a solid threat model which is, ideally, practically motivated.  
Although previous surveys in model stealing~\cite{oliynyk_i_2023} have made an attempt to characterize model stealing attacks, they focus on the subtleties of different attacker's goals like stealing functionality or model weights, and did not establish threat models. 
As we will see in this paper, the threat models discussed in literature that steal the model's functionality vary greatly. 

For other attacks on machine learning, such as evasion attacks, clear threat models have been established~\cite{biggio_wild_2018}. Based on these threat models, guidelines to evaluate defences~\cite{carlini_evaluating_2019} based on adaptive attacks for adversarial examples~\cite{carlini_evaluating_2019} or backdoors~\cite{tan_bypassing_2020} have been established. Such clear threat models allow to match attacks and defenses~\cite{cina_wild_2023} or generally compare or benchmark attacks within one threat model~\cite{croce_robustbench_2021,wu_backdoorbench_2022}. Such a threat model is, for model stealing, to the best of our knowledge, currently missing; and there are no best practices regarding model stealing evaluations either. 

While all fields within the area of ML security have received criticism for not evaluating on settings that are practical enough to be realistic~\cite{apruzzese_real_2023,grosse_towards_2024}, previous work has measured accessibility and granted queries required for model stealing attacks~\cite{grosse_towards_2024} and shown that model stealing is perceived as a relevant threat by practitioners~\cite{grosse_machine_2023}. However, neither work surveys a precise threat model for model stealing, be it practical or not, as it occurs in academic works. We attempt to close this gap.

Orthogonal to our work, but listed for completeness, are approaches that identify wrong configurations, for example for evasion attacks~\cite{pintor_indicators_2022}. Finding such failure cases for model stealing is beyond the scope of this work but would be, alongside benchmarks~\cite{croce_robustbench_2021,wu_backdoorbench_2022}, highly useful for the field of model stealing.

\section{Conclusion} \label{sec:conclusion}
Development of standardised evaluation methodology enables fair comparison among prior works, clear assessment of the current status of the research field, and, what is especially important in case of adversarial machine learning, facilitates countermeasure development. This work presents the first systematic effort for developing such methodology in the field of model stealing. We propose the first comprehensive threat model, and build the first attack comparison framework based on it. By extensively analysing prior attacks on image classification models, we derive best practical recommendations for designing, conducting, and evaluating model stealing attacks. Finally, we raise an in-depth set of research questions concerning evaluation methodology of model stealing attacks.

\begin{acks}
The financial support by the Austrian Federal Ministry of Economy, Energy and Tourism, the National Foundation for Research, Technology and Development, the Christian Doppler Research Association and SBA Research (SBA-K1 NGC), a COMET Centre within the COMET – Competence Centers for Excellent Technologies Programme funded by BMIMI, BMWET, and the state of Vienna, managed by FFG, is gratefully acknowledged. This work has also received funding from the European Union’s Horizon Europe Research and Innovation Programme under grant agreement No 101136305.
Partner Semmelweis University received funding from the Hungarian National Research, Development and Innovation Fund.
\end{acks}

\balance

\bibliographystyle{ACM-Reference-Format}
\bibliography{references}

\appendix

\newpage
\section{Experiment Setup per Paper}
\label{app:setup}

\Cref{tab:exp_setup_per_paper} presents experiment setups from papers listed in \Cref{tab:threat_model}.

\begin{table*}[ht!]
\centering
\footnotesize
\caption{Original datasets, target and substitute architectures per paper.}
\label{tab:exp_setup_per_paper}
\resizebox{\textwidth}{!}{
\begin{tabular}{p{0.05\linewidth}p{0.31\linewidth}p{0.32\linewidth}p{0.32\linewidth}}
\toprule
\multirow{3}{*}{\textbf{Paper}}       & \multirow{3}{*}{\textbf{Original dataset}}              & \multirow{3}{*}{\textbf{Target architecture}}                                                                                                                                             & \multirow{3}{*}{\textbf{Substitute architecture}}                                                                                                                                         \\
                                      &                                                         &                                                                                                                                                                                           &                                                                                                                                                                                           \\
                                      &                                                         &                                                                                                                                                                                           &                                                                                                                                                                                           \\ \midrule
\cite{jagielski_high_2020}            & CIFAR-10, SVHN                        & WSL, WideResNet-28-2                  & ResNet-v2-50, ResNet-v2-200, \unknown \\
\cite{papernot_practical_2017}        & GTSRB, MNIST                                            & \unknown                                                                                                                                                                                  & Custom                                                                                                                                                                                    \\ 
\cite{correia-silva_copycat_2018}     & AR Face, BU3DFE,  JAFFE, MMI, RaFD, CIFAR-10            & \unknown                                                                                                                                                                                  & VGG-16                                                                                                                                                                                    \\ 
\cite{pal_activethief_2020}           & CIFAR10, GTSRB, MNIST                                   & Custom                                                                                                                                                                                    & Custom                                                                                                                                                                                    \\ 
\cite{juuti_prada_2019}               & GTSRB, MNIST                                            & Custom                                                                                                                                                                                    & Custom                                                                                                                                                                                    \\ 
\cite{orekondy_knockoff_2019}         & Caltech256, CUB-200-2011, Diabetic5, Indoor67           & ResNet-34, VGG-16                                                                                                                                                                         & AlexNet, DenseNet-161, ResNet-18, ResNet-34, ResNet-50, VGG-16                                                                                                                            \\ 
\cite{atli_extraction_2020}           & Caltech256, CIFAR-10, CUB-200-2011, Diabetic5, Indoor67 & ResNet-34                                                                                                                                                                                 & ResNet-34                                                                                                                                                                                 \\ 
\cite{pengcheng_query-efficient_2018} & CIFAR-10, FMNIST, MNIST                                 & \unknown, some ResNet                                                                                                                                                                     & Custom, VGG-19                                                                                                                                                                            \\ 
\cite{pal_framework_2019}             & CIFAR-10, FMNIST, GTSRB, MNIST                          & Custom                                                                                                                                                                                    & Custom                                                                                                                                                                                    \\ 
\cite{mosafi_stealing_2019}           & CIFAR-10                                                & Custom                                                                                                                                                                                    & VGG-16                                                                                                                                                                                    \\ 
\cite{yuan_es_2022}                   & CIFAR-10, KMNIST, MNIST, SVHN                           & LeNet, ResNet-18, ResNet-34                                                                                                                                                               & LeNet, ResNet-18, ResNet-34                                                                                                                                                               \\ 
\cite{milli_model_2019}               & CIFAR-10, MNIST                                         & Custom, MLogReg, ResNet-18, VGG-11                                                                                                                                                        & Custom, MLogReg, ResNet-18, VGG-11                                                                                                                                                        \\ 
\cite{kariyappa_maze_2021}            & CIFAR-10, FMNIST, GTSRB, SVHN                           & LeNet, ResNet-20                                                                                                                                                                          & WideResNet-22                                                                                                                                                                             \\ 
\cite{roberts_model_2019}             & FMNIST, KMNIST, MNIST, notMNIST                         & Custom                                                                                                                                                                                    & Custom                                                                                                                                                                                    \\ 
\cite{barbalau_black-box_2020}        & CIFAR-10, FMNIST, 10 Monkey Species                     & AlexNet, LeNet, ResNet-18, VGG-16                                                                                                                                                         & half-AlexNet, half-LeNet, ResNet-18, VGG-16                                                                                                                                               \\ 
\cite{yu_cloudleak_2020}              & GTSRB, VGG Flower                                       & AlexNet, ResNet-50, VGG-19, VGG-Face                                                                                                                                                      & ResNet-50, VGG-19, VGG-Face                                                                                                                                                               \\ 
\cite{gong_inversenet_2021}           & CIFAR-10, GTSRB, MNIST                                  & Custom                                                                                                                                                                                    & Custom, ResNet-18                                                                                                                                                                         \\ 
\cite{truong_data-free_2021}          & CIFAR-10, SVHN                                          & ResNet-34                                                                                                                                                                                 & ResNet-18                                                                                                                                                                                 \\ 
\cite{miura_megex_2021}               & CIFAR-10, SVHN                                          & ResNet-34                                                                                                                                                                                 & ResNet-18                                                                                                                                                                                 \\ 
\cite{sanyal_towards_2022}            & CIFAR-10, CIFAR-100                                     & AlexNet, ResNet-18, ResNet-34                                                                                                                                                             & half-AlexNet, ResNet-18                                                                                                                                                                   \\ 
\cite{zhang_thief_2021}               & Caltech256, ImageNet, FMNIST                                                  & AlexNet, Custom, LeNet, ResNet-34, ResNet-50                                                                                                                                                                    & AlexNet, Custom, LeNet, ResNet-18, ResNet-34, ResNet-50                                                                                                                                                                  \\ 
\cite{wang_enhance_2022}              & Caltech256, CIFAR-10, CUB-200-2011, SVHN                & ResNet-34                                                                                                                                                                                 & \unknown                                                                                                                                                                                  \\ 
\cite{wang_black-box_2022}            & Caltech256, CIFAR-10, CUB-200-2011, SVHN                & ResNet-34                                                                                                                                                                                 & some DenseNet, ResNet-18, ResNet-34, ResNet-50, VGG-16                                                                                                                                    \\ 
\cite{yan_towards_2022}               & CIFAR-10, CIFAR-100, FMNIST, MNIST                      & DenseNet-161, ResNet-50, VGG-19                                                                                                                                                           & Custom                                                                                                                                                                                    \\ 
\cite{xie_game_2022}                  & BelgiumTSC, MNIST                                       & AlexNet, LeNet                                                                                                                                                                            & half-AlexNet, half-LeNet, ResNet-18, VGG-16                                                                                                                                               \\ 
\cite{chen_d-dae_2023}                & CIFAR-10, FMNIST, GTSRB, ImageNette, MNIST              & LeNet, ResNet-34, VGG-16                                                                                                                                                                  & LeNet, ResNet-34, VGG-16                                                                                                                                                                  \\ 
\cite{he_drmi_2021}                   & CIFAR-10, ImageNet, MNIST                               & Custom, Inception-V3, LeNet, ResNet-18, ResNet-152                                                                                                                                        & Custom, Inception-V3, LeNet, ResNet-18, ResNet-152                                                                                                                                        \\ 
\cite{liu_ml-doctor_2022}             & CelebA, FMNIST, STL-10, UTKFace                         & AlexNet, Custom, ResNet-18, VGG-19, Xception                                                                                                                                              & AlexNet, Custom, ResNet-18, VGG-19, Xception                                                                                                                                              \\ 
\cite{rosenthal_disguide_2023}        & CIFAR-10, CIFAR-100                                     & ResNet-18, ResNet-34                                                                                                                                                                      & ResNet-18                                                                                                                                                                                 \\ 
\cite{yan_holistic_2023}              & CIFAR-10, CIFAR-100, FMNIST, MNIST                      & DenseNet-121, DenseNet-161, DenseNet-169, DenseNet-201, Inception-V1, Inception-V2, Inception-V3, ResNet-18, ResNet-34, ResNet-50, ResNet-101, ResNet-152, VGG-11, VGG-13, VGG-16, VGG-19 & DenseNet-121, DenseNet-161, DenseNet-169, DenseNet-201, Inception-V1, Inception-V2, Inception-V3, ResNet-18, ResNet-34, ResNet-50, ResNet-101, ResNet-152, VGG-11, VGG-13, VGG-16, VGG-19 \\ 
\cite{zhang_towards_2022}             & CIFAR-10, CIFAR-100, FMNIST, MNIST, SVHN, TinyImageNet  & \unknown, Custom, ResNet-34                                                                                                                                                               & \unknown, Custom, ResNet-18                                                                                                                                                               \\ 
\cite{yan_explanation-based_2023}     & CIFAR-10, SVHN                                          & ResNet-34                                                                                                                                                                                 & ResNet-18                                                                                                                                                                                 \\ 
\cite{yan_explanation_2023}           & CIFAR-10, CIFAR-100, FMNIST, MNIST                      & DenseNet-161, ResNet-50, VGG-19                                                                                                                                                           & \unknown                                                                                                                                                                                  \\ 
\cite{yang_efficient_2023}            & CIFAR-10, FMNIST, MNIST                                 & AlexNet, LeNet                                                                                                                                                                            & half-AlexNet, half-LeNet                                                                                                                                                                  \\ 
\cite{lin_quda_2023}                  & CIFAR-10, FMNIST                                        & AlexNet, LeNet, ResNet-18, VGG-11                                                                                                                                                         & AlexNet, ResNet-18, VGG-11                                                                                                                                                                \\ 
\cite{liu_shrewdattack_2023}          & FMNIST, Intel-Image                                     & \unknown                                                                                                                                                                                  & SqueezeNet                                                                                                                                                                                \\ 
\cite{pape_limitations_2023}          & CIFAR-10, SVHN                                          & \unknown, ResNet-152                                                                                                                                                                      & Custom, Inception-V3, ResNet-152                                                                                                                                                          \\ 
\cite{liu_efficient_2024}            & CelebA, CIFAR-10, SVHN                                  & ResNet-34                                                                                                                                                                                 & ResNet-18                                                                                                                                                                                 \\ 
\cite{hondru_towards_2025}            & CIFAR-10, Food-101                                      & AlexNet, ResNet-50                                                                                                                                                                        & half-AlexNet, ResNet-18                                                                                                                                                                   \\ 
\cite{khaled_careful_2022}            & CIFAR-10, MNIST, SVHN                                   & Custom, ResNet-34                                                                                                                                                                         & LeNet, VGG-16                                                                                                                                                                             \\ 
\cite{zhao_extracting_2023}           & CIFAR-10, Flower-17, GTSRB, STL-10                      & VGG-13                                                                                                                                                                                    & ResNet-50                                                                                                                                                                                 \\ 
\cite{karmakar_marich_2023}           & CIFAR-10, MNIST                                         & Custom, some ResNet                                                                                                                                                                       & Custom, ResNet-18                                                                                                                                                                         \\ 
\cite{jindal_army_2024}        & Caltech256, CIFAR-10, CIFAR-100, CUB-200-2011           & ResNet-34                                                                                                                                                                                 & AlexNet, DenseNet-121, EfficientNet-B2, MobileNet-V3, ResNet-34                                                                                                                           \\ 
\cite{beetham_dual_2023}              & CIFAR-10, CIFAR-100, FMNIST, GTSRB, MNIST, SVHN         & ResNet-34                                                                                                                                                                                 & ResNet-18                                                                                                                                                                                 \\ 
\cite{hong_exploring_2023}            & CIFAR-10, FMNIST, MNIST, SVHN                           & ResNet-34, VGG-16                                                                                                                                                                         & VGG-11                                                                                                                                                                                    \\ 
\cite{yang_swifttheft_2024}         & CIFAR-10, GTSRB, VGG Flower                             & Custom                                                                                                                                                                                    & Custom                                                                                                                                                                                    \\ 
\cite{li_model_2025}                 & CIFAR-10, FMNIST, GTSRB, SVHN                           & some MobileNet, some ResNet, some VGG                                                                                                                                                     & some MobileNet, some ResNet, some VGG                                                                                                                                                     \\ \bottomrule
\end{tabular}

}
\end{table*}

\end{document}